\documentclass[conference]{IEEEtran}
\IEEEoverridecommandlockouts

\usepackage{amsmath,amssymb,amsfonts}
\usepackage{mathtools}
\usepackage[linesnumbered,ruled]{algorithm2e}
\usepackage{algpseudocode} 
\usepackage{subfigure}
\usepackage{enumitem}

\usepackage{threeparttable}
\usepackage{booktabs}
\usepackage{makecell}  
\usepackage{arydshln} 
\usepackage{multirow}
\usepackage{adjustbox}
\usepackage{stackengine}

\usepackage[dvipsnames]{xcolor}
\definecolor{background}{HTML}{FFFFF5}
\definecolor{edge}{HTML}{87C2B1}
\definecolor{bisque}{HTML}{B86500}
\definecolor{myblue}{HTML}{2A4597}

\usepackage[colorlinks=true,citecolor=black, linkcolor=black, urlcolor=myblue]{hyperref}
\usepackage{cleveref}

\usepackage[most]{tcolorbox}

\definecolor{beauty}{HTML}{58A0A4}
\definecolor{myblue}{HTML}{2A4597}
\definecolor{luoshen}{HTML}{C5563B}
\definecolor{background}{HTML}{D0EEF3}
\definecolor{edge}{HTML}{6DAECD}

\newtcolorbox{mybox}{
colback=background!20, colframe=edge,
width=\columnwidth,% total width
arc=1.5mm,
auto
outer
arc
}

\usepackage[]{collab}

\usepackage{xspace}
\def\tool{{SPES}\xspace}

\begin{document}
\title{\tool: Towards Optimizing Performance-Resource Trade-Off for Serverless Functions}

\author{
  \IEEEauthorblockN{
    Cheryl Lee\IEEEauthorrefmark{1},
    Zhouruixing Zhu\IEEEauthorrefmark{2},
    Tianyi Yang\IEEEauthorrefmark{1},
    Yintong Huo\IEEEauthorrefmark{1},
    Yuxin Su\IEEEauthorrefmark{3}, 
    Pinjia He\IEEEauthorrefmark{2},
    and
    Michael R. Lyu\IEEEauthorrefmark{1}
  }

  \IEEEauthorblockA{\IEEEauthorrefmark{1}The Chinese University of Hong Kong, Hong Kong, China. \\
    Email: cheryllee@link.cuhk.edu.hk, \{tyyang, ythuo, lyu\}@cse.cuhk.edu.hk}

    \IEEEauthorblockA{\IEEEauthorrefmark{2}The Chinese University of Hong Kong, Shenzhen, Shenzhen, China. \\
    Email: zhouruixingzhu@link.cuhk.edu.cn, hepinjia@cuhk.edu.cn}

  \IEEEauthorblockA{\IEEEauthorrefmark{3}School of Software Engineering, Sun Yat-sen University, Zhuhai, China. Email: suyx35@mail.sysu.edu.cn}
}

\maketitle

\begin{abstract}
As an emerging cloud computing deployment paradigm, serverless computing is gaining traction due to its efficiency and ability to harness on-demand cloud resources. 
However, a significant hurdle remains in the form of the \textit{cold start} problem, causing latency when launching new function instances from scratch.
Existing solutions tend to use over-simplistic strategies for function pre-loading/unloading without full invocation pattern exploitation, rendering unsatisfactory optimization of the trade-off between cold start latency and resource waste.
To bridge this gap, we propose \textbf{\tool}, the first differentiated scheduler for runtime cold start mitigation by optimizing serverless function provision.
Our insight is that the common architecture of serverless systems prompts the concentration of certain invocation patterns, leading to predictable invocation behaviors.
This allows us to categorize functions and pre-load/unload proper function instances with finer-grained strategies based on accurate invocation prediction.
Experiments demonstrate the success of \tool in optimizing serverless function provision on both sides: reducing the $75^{th}$-percentile cold start rates by 49.77\% and the wasted memory time by 56.43\%, compared to the state-of-the-art.
By mitigating the cold start issue, \tool is a promising advancement in facilitating cloud services deployed on serverless architectures.
\end{abstract}

\begin{IEEEkeywords}
Cloud computing, serverless computing, cold start, function categorization
\end{IEEEkeywords}

\section{INTRODUCTION}
\begin{figure}[htb]
    \centering
    \vspace{-0.1in}
        {\includegraphics[width=0.98\linewidth]{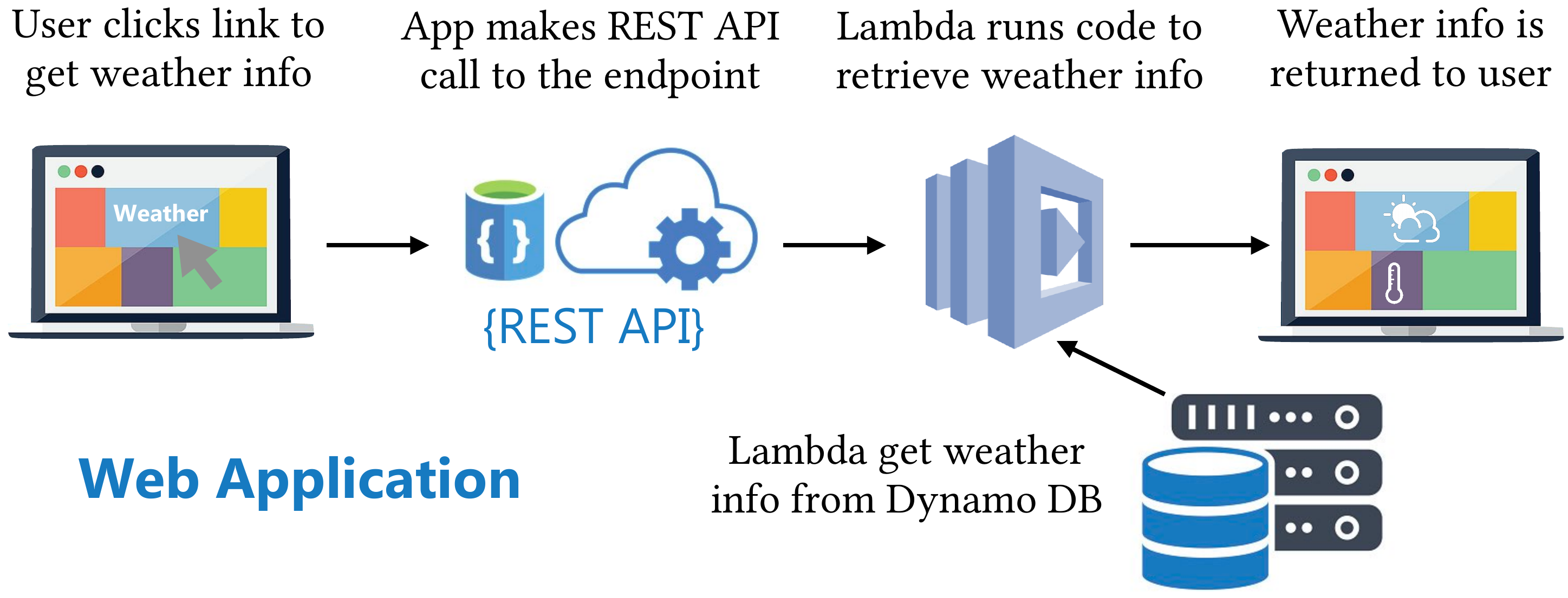}}
        \vspace{-0.1in}
    \caption{An weather inquiry website based on a serverless web application.}
    \vspace{-0.2in}
    \label{fig:intro_example}
\end{figure}

Function-as-a-service (FaaS), the most prominent implementation pattern of serverless computing, has extensively simplified developers' access to cloud resources.
Major cloud vendors such as AWS Lambda, Google Cloud Functions, and Azure Functions have supported FaaS-based web services~\cite{WebServerless, Firestore}, machine learning~\cite{INFless, Cirrus}, and other cloud applications~\cite{Freyr, ServerlessBFT}.
FaaS shifts the burden of infrastructure management from developers to cloud vendors, allowing developers to focus on their application functionality~\cite{Valve}. 
The statelessness and event-driven architecture of FaaS facilitate seamless updates and flexible application deployment.
Plus, FaaS applies a pay-as-you-go billing model and dynamically allocates resources based on demand~\cite{FaaSnap}, appealing to users via cost-saving benefits.
Figure~\ref{fig:intro_example} depicts a serverless application example.

Despite their benefits, cloud users on serverless platforms suffer from the infamous cold-start problem, which causes them to trade off performance latency and memory costs.
Serverless functions are spawned on instances and typically have very short execution duration (orders of milliseconds to seconds)~\cite{Formalfoundations}.
However, booting a function from scratch (\textit{i.e., cold start}) incurs expensive latency in preparing the execution environment~\cite{Firecracker}, compared to running an \textit{warm} function whose instance is already loaded in the memory.
Cold-start latency can account for 80\% of the total response latency~\cite{StaticAnalysis}, leading to user dissatisfaction and eventually causing user churn and economic loss.
The pay-per-use model of FaaS also motivates cloud vendors to avoid unbillable set-up latency~\cite{Economics}.
Meanwhile, keeping all functions warm is infeasible due to two reasons.
First, most functions are invoked infrequently~\cite{ServerlessWild}.
It is unacceptable to keep idle (also unbillable) and infrequently used function instances in memory, as it leads to unnecessary costs and wastes valuable resources.
Second, because of the infrequent invocation and short execution, FaaS providers colocate thousands of function instances on a single server to achieve full utilization~\cite{vHive}. 
Loading all functions can occupy hundreds of GBs of memory, far beyond a server's capacity~\cite{vHive}. 
Thus, reducing cold start latency while minimizing memory waste has been a key challenge~\cite{FaasCache, INFless, SCache}.

Existing solutions generally fall into two categories~\cite{currentTrents,SOCK}: accelerating the setup~\cite{Catalyzer, FaaSFlow, FaaSLight} or reducing cold start occurrences~\cite{Defuse, ServerlessWild, Ensure}.
This paper focuses on the reduction of runtime cold start occurrence without renovating the underlying system-layer infrastructure.
Previous such efforts~\cite{Defuse, ServerlessWild, Ensure} either keep instances loaded for a fixed period post-execution or manage instances based on rudimentary inter-invocation intervals.
However, they overlook the intrinsic influence of triggering events on invocation patterns, thus falling short in addressing infrequently invoked functions.
Hence, we intend to develop a more effective, non-intrusive method that can load functions right before incoming invocations and recycle idle instances with no potential near-future invocations.
The key is to predict invocations accurately.

We identify four challenges in developing such a method:
\begin{itemize}[leftmargin=12pt, topsep=0pt]

\item \textit{Efficiency}: A FaaS platform can receive thousands of function invocations every minute from millions of potential functions~\cite{ServerlessData}. Most functions have a very short execution time, so the method must decide the provision within a limited and unbillable time.

\item \textit{Scalability}: The method should be highly elastic, quickly scaling up and down to meet invocation fluctuations promptly, as the requests of FaaS functions are usually bursty and dynamic~\cite{Firestore}.

\item \textit{Imbalance}: The distribution of function invocations can be highly imbalanced, and most functions are rarely invoked~\cite{ServerlessWild}. This poses a challenge to approaches requiring sufficient training data.

\item \textit{Evolution}: The lightweight nature of FaaS makes applications evolve faster than traditional software. Both short-term factors (e.g., user activity variations) and long-term factors (e.g., software updates) can cause concept shifts in function invocations, thereby hindering predictive models.

\end{itemize}

\textbf{Our insight:} \textit{the event-driven architecture of serverless systems enables identifying common invocation patterns, allowing for efficient function categorization and provision with bespoke strategies.}
This insight is grounded in both domain expertise and data analysis.
Serverless functions are triggered by business-specific events and often fall into limited types, laying a foundation for invocation prediction.
For example, the invocations of a scheduled data processing function are easy to predict, and so is the pre-loading.
While provision is trigger-agnostic, invocation patterns can be modeled from historical invocations associated with triggers.
Moreover, functions within an application often display behavior convergence. For instance, functions within a multi-stage processing workload are typically invoked in turn.
Our empirical analysis also reveals such consistent behavior patterns.
Besides, some functions with irregular invocations may exhibit temporal locality, making it feasible to keep them loaded temporarily, reducing cold starts with minimal memory overhead.
Thus, our study shows that functions can present limited predictable patterns, allowing for differentiated strategic pre-loading to minimize cold starts with minimal memory impact.

Hence, we propose \textbf{\tool}, a differentiated \underline{S}cheduler for \underline{P}rovisioning runtim\underline{E} \underline{S}erverless functions in order to optimize the trade-off between performance latency and resource cost.
After extensive data analysis, we summarize five typical invocation patterns and design three extra policies for functions experiencing less predictive invocations.
We also propose a simple but effective metric, \textit{i.e.} co-occurrence rate, to measure the correlation degree between functions. Highly correlated functions can be predictive indicators of each other.
In addition, we design adaptive strategies to handle the concept shifts as FaaS evolves.
Hence, \tool can predict the next invocation accurately and provision the function in advance with fine-grained strategies.
As a result, our rule-based \tool addresses all the above challenges. Its tailored strategies can handle diverse functions with remarkable scalability and adaptivity, introducing a small computation overhead. 

We evaluate \tool on the most widely used industrial dataset released by Microsoft Azure~\cite{ServerlessWild}.
Experiments demonstrate the superiority of \tool, which achieves both-side improvement:
reducing the $75^{th}$-percentile cold start rate by 49.77\%--89.20\% and the wasted memory time by 13.51\%–-64.58\% compared to baselines, and also reducing the proportion of ``always-cold'' functions significantly.
Extensive ablation studies further confirm the effectiveness of our designs of inter-function connection and adaptive strategies.

In summary, the main contributions of this paper are:
\begin{itemize}[leftmargin=12pt, topsep=0pt]
\item We propose \tool, the first differentiated method to mitigate the cold start issue by optimizing the runtime provision of serverless functions based on invocation prediction. 

\item We summarize typical invocation patterns and categorize functions accordingly as the basis of tailored provision strategies. 
We also create an effective ``co-occurrence'' metric to gauge inter-function correlations.
Our code is available at~\textcolor{myblue}{\url{https://github.com/BEbillionaireUSD/SPES}}.

\item We conduct extensive experiments to demonstrate the significant improvement of \tool in the cold start and wasted memory reduction, as well as the contribution of designs for adaptivity and unseen functions.

\end{itemize}

\section{Background}\label{sec:background}
\subsection{Serverless Computing}
Serverless computing represents an emerging cloud programming paradigm where cloud providers fully manage the underlying infrastructure and allocate resources dynamically, enabling developers to concentrate solely on their application's core logic. 
In FaaS, developers implement services as stateless workloads, \textit{i.e.} functions, designed to respond to an individual event using pre-defined rules, called the \textit{trigger}~\cite{Netherite}.
Developers only pay for the actual computation resources on a per-execution basis, making it cost-efficient and scalable~\cite{ServerlessDataScience}. 
Besides, a could service-based \textit{application} is usually broken down into separate, independent functions in practical implementation, where these functions are sometimes explicitly chained together~\cite{Kraken}.
The adoption of serverless computing has witnessed rapid growth, especially in web services, machine learning training~\cite{TowardsML}, etc.
Notwithstanding its advantages, serverless computing poses new challenges, such as ensuring that a service adheres to quality of service (QoS) demands related to response time or tail latency~\cite{AQUATOPE}.

\subsection{Cold Start Challenge}\label{sec:background:challenge}
\begin{figure}[htb]
    \centering
     \vspace{-0.12in}
        {\includegraphics[width=0.98\linewidth]{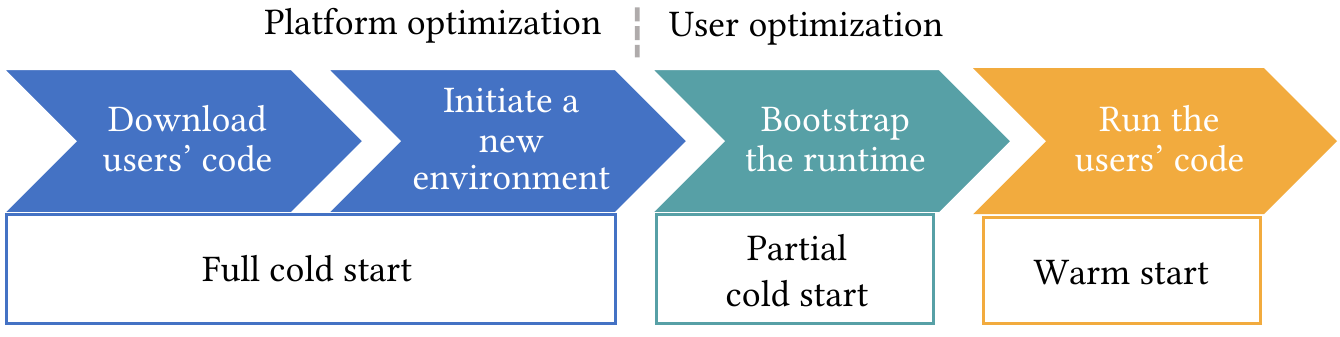}}
     \vspace{-0.1in}
    \caption{A serverless function's lifecycle.}
     \vspace{-0.12in}
    \label{fig:function_process}
\end{figure}

Figure~\ref{fig:function_process} displays a serverless function's lifecycle. 
A FaaS provider first downloads the code of users and initiates the execution environment (or sandbox) in the memory on cluster machines, known as a \textit{cold start}. 
A cold start involves retrieving code from storage, container (or VM) initialization, loading code into memory, and executing the function's handler. In contrast, a \textit{warm} start jumps directly to execution, resulting in a faster response.
Upon serving a request, its environment remains idle for a while without other invocations before the orchestration system decommissions it~\cite{Atoll}.
Naturally, the more reused environments, the less cold starts.
There exists a trade-off in keeping the instance alive: saving start-up resources and speeding up subsequent requests but incurring \textit{idle time} costs. 
Such costs can be quantified by \textit{wasted memory time (WMT)}, i.e., the time when the image of a function is kept in memory, but the function is not actually invoked. WMT is an important metric to measure resource waste in practice~\cite{ServerlessWild}.

The cold start problem is critical because it significantly increases runtime latency, yet serverless providers should meet the QoS requirements and maintain an ``always-ready'' illusion to users.
On the one hand, cold starts can dominate the overall execution time~\cite{FaaSLight, FaasCache, Peeking}. \cite{Catalyzer} shows that the ``Execution/Overall'' latency ratio of most tested functions in gVisor can not even achieve 30\%.
On the other hand, the cold start happens very frequently~\cite{AJAX, SOCK, FaaSNet}.
Hence, the additional latency incurred by cold starts is exceptionally unbearable.
Considering that memory is limited and expensive, mitigating the cold start challenge aims to either load/unload proper function instances or speed up cold start initiation.

\section{Preliminary Empirical Analysis}\label{sec:motivation}
\subsection{Challenges of Serverless Function Provision}\label{sec:motivation:challenge}

\subsubsection{Efficiency requirement}
Serverless provision must make immediate decisions for massive invocations, which hinders the potential of applying complex models.
This requirement stems from function massiveness and latency sensitivity.
Large-scale FaaS platforms launch a burst-parallel swarm and allow for thousands of invocations every minute~\cite{LaptopLambda}, all waiting for the provision decision.
Moreover, serverless functions are usually short-lived, so a provisioning method must handle invocations and update its strategy quickly during the unbillable scheduling time.
As studied in~\cite{Hermod}, it is better to schedule invocations for execution as soon as they enter the system. 
A fast and simple provision method is thereby much more desired.

\subsubsection{Scalability under invocation spikes}
The provisioning method is also expected to scale automatically to a very high load.
FaaS invocations are extremely bursty and dynamic~\cite{FaaSNet}.
An ideal FaaS platform should handle the invocation burst, so the provision decision-maker must be capable of rapid and elastic scaling to respond to sudden invocation spikes seamlessly and scale down during periods of low demand.
For example, during a holiday shopping period, an e-commerce site running on a FaaS architecture may experience a 10x increase in traffic and function invocations compared to a normal workday. 
Hence, we require a highly scalable solution.

\subsubsection{Imbalance in invocation distribution}
\begin{figure}[htb]
    \centering
     \vspace{-0.1in}
        {\includegraphics[width=0.98\linewidth]{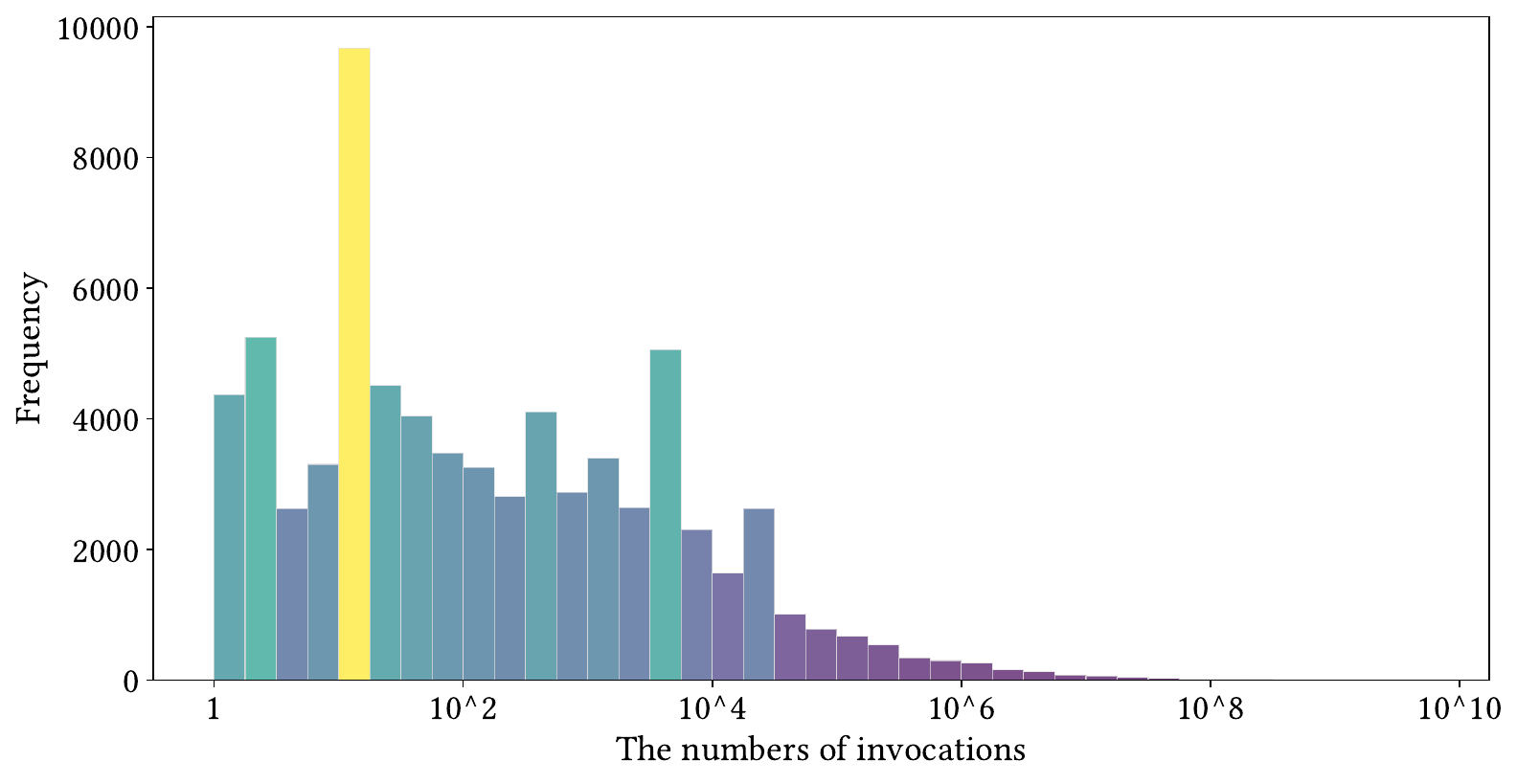}}
         \vspace{-0.1in}
    \caption{The distribution of function invocations. The x-axis represents the range of function invocation counts, and the y-axis indicates the number of functions falling into the corresponding ranges.}
     \vspace{-0.1in}
    \label{fig:lognum_dist}
\end{figure}

Serverless functions exhibit heterogeneity in functionality, manifesting distinct implementations through a diversity of libraries and runtimes~\cite{AQUATOPE}. 
Moreover, the functions are triggered in response to various events, such as HTTP requests, data updates, and the receipt of notifications~\cite{RiseServerless}. 
The various real-world demands make the frequencies of function invocations dramatically different.
Figure~\ref{fig:lognum_dist} shows the distribution of invocations from an industrial dataset released by Microsoft Azure~\cite{ServerlessWild}, collected for 14 days.
We can observe that the invocations are highly non-uniform, with most functions being rarely invoked.
This imbalance poses a challenge to learning-based approaches that require lots of records for training because many functions may not have invocation records in the data collection.

\subsubsection{Evolution in invocation behavior}\label{sec:motivation:challenges:evolution}
\begin{figure}[htb]
    \centering
     \vspace{-0.1in}
        {\includegraphics[width=0.95\linewidth]{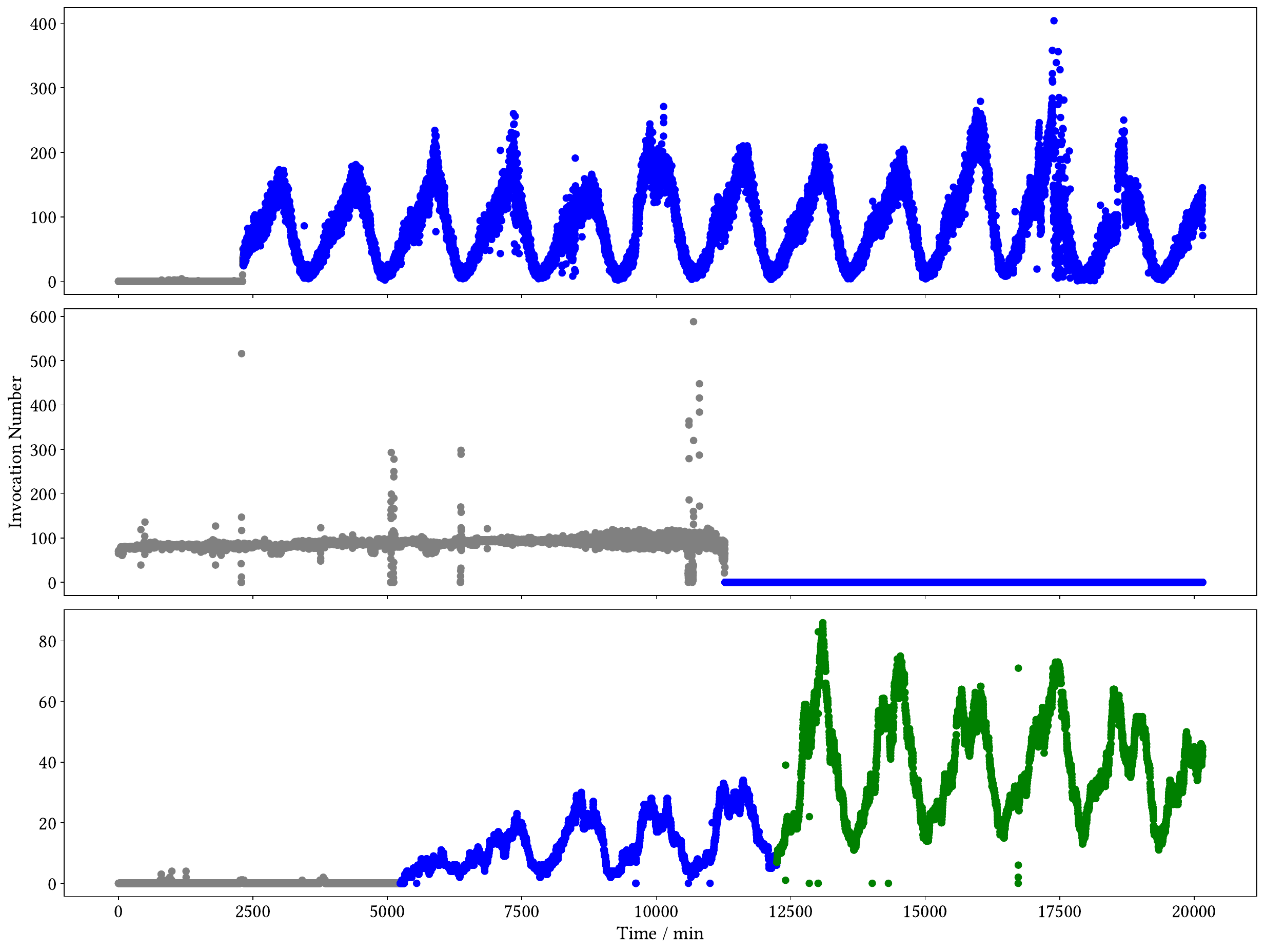}}
         \vspace{-0.1in}
    \caption{The function invocations can experience distinct concept shifts that may degrade the provision performance. Different colors distinguish changes over time in the function invocation patterns.}
     \vspace{-0.1in}
    \label{fig:show_shift}
\end{figure}

Invocation evolution can degrade the performance of provisioning strategies because they rely on proactively pre-loading to-be-invoked functions to reduce cold starts fundamentally, whose predictive modeling can result in poor performance with invocation evolution.
The situation is even worse for learning-based approaches, which may require frequent, time-consuming re-training.
Many factors can cause such shifts, \textit{e.g.} seasonal traffic, data volume changes, geographic expansion, etc.
Considering the e-commerce site mentioned earlier, it can experience short-term spikes during holiday seasons and evolve to long-term increases when expanding to overseas markets. 
Changes in business logic are eventually propagated to the function development and utilization, resulting in observable shifts in invocations.
In the scope of a single serverless function, a function can be re-bound to a new trigger or new calling chains, leading to noticeable concept shifts.
We also observe the short-term evolution of invocations in the mentioned dataset, as shown in Figure~\ref{fig:show_shift}, which plots the invocations of three typical functions over time.
This fast-evolving characteristic of FaaS requires an adaptive provision method capable of handling concept shifts.

\subsection{Observations and Our Insight}\label{sec:motivation:insight}
\subsubsection{Invocation pattern and triggers}\label{sec:motivation:insight:trigger}
\begin{figure}[htb]
    \centering
    \vspace{-0.2in}
        {\includegraphics[width=0.95\linewidth]{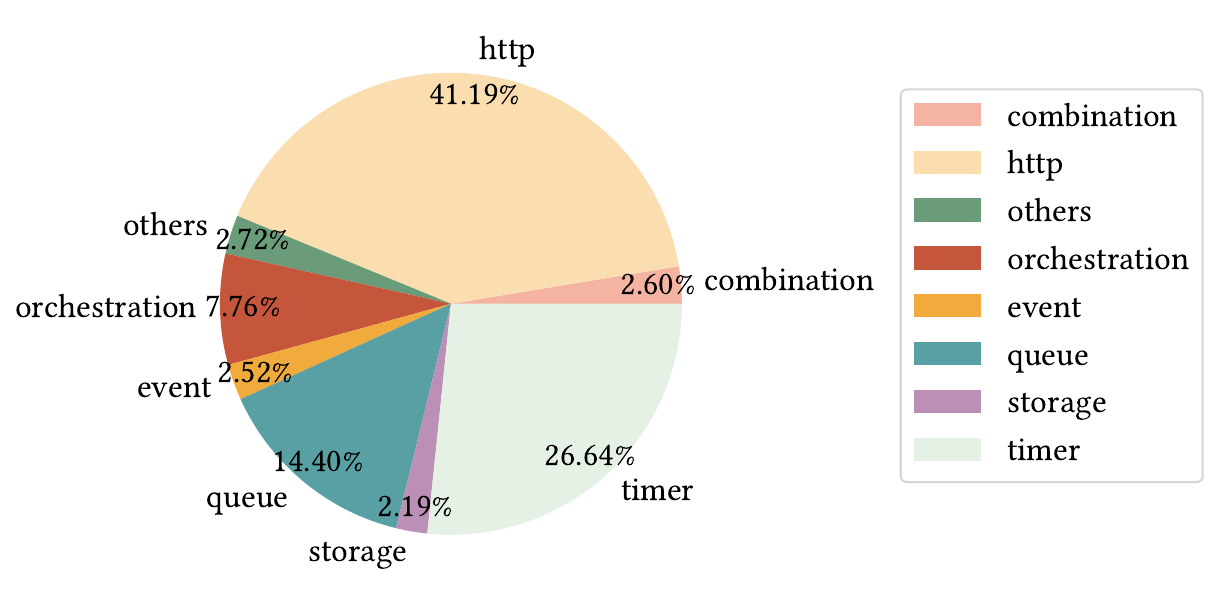}}
     \vspace{-0.1in}
    \caption{The proportion of trigger types among functions.}
     \vspace{-0.1in}
    \label{fig:trigger_prop} 
\end{figure}

Serverless systems operate on an event-driven architecture. Generally, a trigger defines how a function is invoked by causing a function to run, so function invocation patterns are decided by pre-defined trigger types, such as HTTP requests, queued messages, and scheduled timers, resulting in predictive behaviors.
Two observations from the mentioned dataset support this argument:

First, most functions correspond to only one trigger type.
The proportion of triggers bound with functions is shown in Figure~\ref{fig:trigger_prop}, where ``combination'' means more than one (at most three) type of triggers bound with one function.
Since only 2.6\% functions have more than one trigger type, we can link each function to one pattern without concerning too much about the derived combination of patterns.

Second, some triggers lead to regular invocation behaviors, offering possible clues for invocation prediction.
We use the Kolmogorov-Smirnov test~\cite{KStest} to check if a single function's invocations with more than 10 counts follow a given distribution ($pValue \geq 0.05$, \textit{i.e.} not rejecting the null hypothesis).
Results show that 68.12\% timer-triggered functions are invoked periodically or quasi-periodically, excluding 6.65\% do not have enough counts.
Also, 45.02\% HTTP-triggered invocations follow a Poisson arrival process, conforming to the expectation of human-generated activities, excluding 36.20\% functions without enough samples.
Though the Poisson process is memoryless and unsuitable for direct invocation prediction, it still indicates that meaningful invocation behaviors are associated with triggers. 

\subsubsection{Application workflow}\label{sec:motivation:insight:app}
An application contains several functions in a logic workflow~\cite{Kraken}, such as function chaining or callback calling, leading to recurring and interdependent invocation behaviors.
Records also show that invocations from one application/user tend to behave consistently. 
We create a metric \textit{co-occurrence rate (COR)} to measure such consistency. 
For each function, we record the time when it and another function are invoked, divided by its invoked time, as its COR.
The idea is borrowed from linguistics, where co-occurrence is interpreted as indicating semantic proximity. 
For each function with at least another function (\textit{candidate function}) sharing the same application/user, we negatively sample 50 no-application/user-overlapped functions.
The average COR of candidate functions is about 4.6x compared to negative samples, which are 0.2312 and 0.0504, respectively.
Inspired by $\S$\ref{sec:motivation:insight:trigger}, we also find that the invocation behavior convergence is more significant when candidates share the same trigger.
Particularly, the average COR rate of the same-trigger-based candidate functions is 0.2710, whereas it is 0.1307 for different-trigger-based candidates. 

\subsubsection{Temporal locality in invocations}
\begin{figure}[htb]
    \centering
    \vspace{-0.12in}
        {\includegraphics[width=0.95\linewidth]{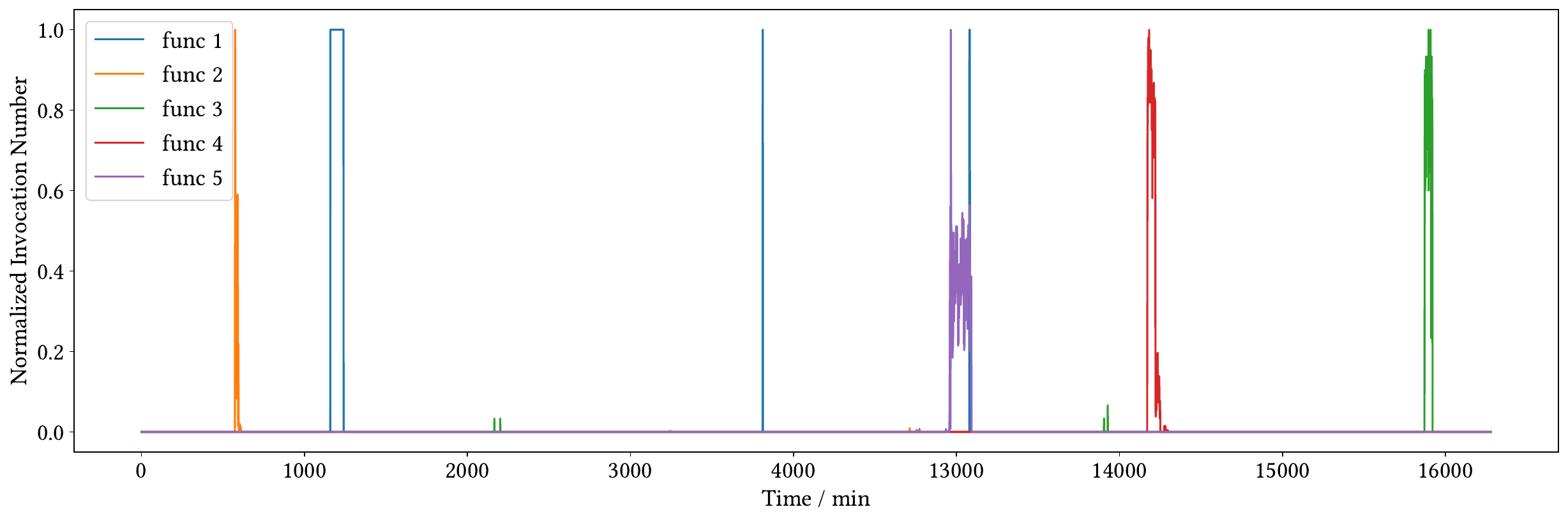}}
        \vspace{-0.1in}
    \caption{Invocations of these infrequently invoked five functions are centralized over several periods. We omit the no-invocation period (4000--13000).}
    \vspace{-0.1in}
    \label{fig:temporal_locality}
\end{figure}

One more noteworthy behavior is temporal locality, the tendency of functions to experience short bursts in a short-term period.
On the one hand, the triggering events can have a rapid succession due to real-world bursty requests (e.g., a high-season e-commerce site).
On the other hand, many common application patterns contribute to temporal locality.
In a fan-out application, a function activates multiple other functions concurrently, often resulting in rapid and successive calls to the secondary functions.
Callback functions that handle asynchronous operations can also undergo multiple invocations in quick succession until the workload is completed.
Such temporal locality also exists in the dataset. Figure~\ref{fig:temporal_locality} plots the minimum-maximum normalized invocations of five infrequently triggered functions, whose invocations are consecutive and concentrated in certain periods.
For infrequently invoked functions presenting temporal locality, invocation prediction can be difficult, yet simply keeping a loaded function for some time can considerably reduce cold starts without wasting much memory.

\vspace{-0.1in}
\begin{mybox}
  \small
  \textbf{Insight}: The microservice and event-driven architecture of serverless systems results in the concentration of recurring and systematic invocation behaviors.
  By analyzing the invocation patterns and categorizing functions, we can accurately predict and pre-load function instances to reduce cold starts.
\end{mybox}

This insight inspires \tool, a rule-based, efficient, and elastic function scheduler.
It leverages the application, user, and trigger information to design tailored and evolution-adaptive strategies to pre-load/evict categorized functions without requiring large amounts of training data.
Overall, \tool provides a language-agnostic, developer-free function provision approach that can be adopted on existing FaaS systems seamlessly to optimize cold starts and memory usage.
\begin{figure*}[htb]
    \centering
    \vspace{-0.1in}
        {\includegraphics[width=0.85\linewidth]{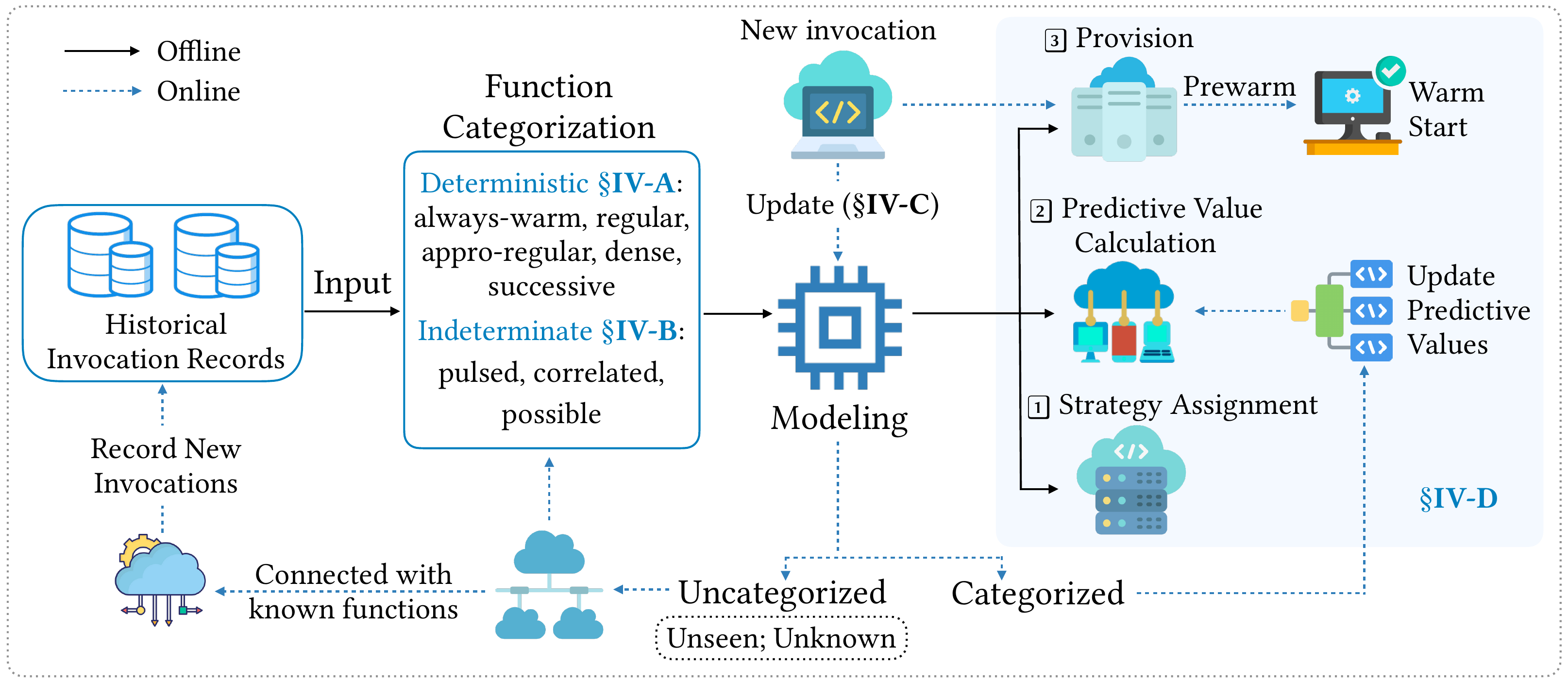}}
        \vspace{-0.1in}
    \caption{Overview of \tool. In the offline phase, \tool models the historical invocation records for each function and categorizes functions according to their mined invocation patterns. During the online phase, if the new invocation comes from a categorized function, its function's provision directly follows its provision rule, and uncategorized functions are tried to be connected with known ones. The invocation prediction and predictive values/indicators are then updated upon new invocations.}
    \vspace{-0.1in}
    \label{fig:overview}
\end{figure*}

\section{METHODOLOGY}\label{sec:method}
This section introduces the design of \tool. 
An ideal scheduler should decide to load a function exactly before its invocation and evict it from memory after the execution if no more invocations are imminent.
The decision-making relies on fine-grained invocation prediction.
To this end, we propose \tool, whose overview is presented in Figure~\ref{fig:overview}.
\tool consists of four parts: deterministic function categorization, indeterminate function assignment, adaptive strategy application, and function provision based on invocation prediction.
Specifically, we first summarize typical invocation patterns, which result from pre-defined triggers~\cite{AzureTriggers} and the combinatorial methods of function calling (in common application scenarios~\cite{AzureScenario}) and then formulate their corresponding definitions.
If a function satisfies the definitions, it is categorized; Otherwise, the indeterminate functions are assigned to three supplementary types.
We also design two extra adaptive strategies to handle the concept shifts as FaaS evolves.
Finally, \tool provisions functions according to rule-based invocation prediction, where each type follows its own predicting rule.

Note that different triggers can exhibit the same timing feature in invocations. For example, the service bus trigger and the event gird trigger both can handle moving datagrams (though focusing on different message types). It is neither necessary nor desirable to correspond each trigger to a function type in the context of the cold start problem.

Let us start with three definitions: 
\begin{itemize}[leftmargin=10pt, topsep=0pt]
    \item \textit{Waiting time (WT)}: the length of successive idle time. 
    Take an invocation sequence $(28, 0, 12, 1, 0, 0, 0, 7)$ as an example, where each value denotes the invocation count at each sampling slot, derived to a sequence of WTs $\{WT\}=(1, 3)$, as no invocation exists at the 2nd slot or in slots 5--8.
    If we foreknow the next WT, for example, at the 4th slot, we know no invocations will arrive in slots 5-8, then we can easily make a perfect decision: evict the function now and re-load it at the end of slot 9.
    Hence, we can transform invocation prediction to WT prediction.
    Different from a previously proposed inter-arrival time (IAT)~\cite{ServerlessWild}, WT depicts the slot-grained interval between two successive invocation sequences, from the last's end to the next's start.
    WT is also different from idle time (IT). IT is the time without invocations or subsequent function executions, so IT exists after any single execution. 
    WT, instead, only appears when a successive series of invocations ends, and the end of a single execution does not necessarily incur a WT.

    \item  \textit{Active time (AT)}: the length of successive time with invocations. 
    The above-mentioned invocation sequence delivers an AT sequence $\{AT\}=(1, 2, 1)$ because the function is invoked in slots 1, 3--4, and 9. 
    
    \item \textit{Active number (AN)}: a descendable definition, the number of successive invocations during AT. Again, the above sequence delivers $\{AN\}=(28, 13, 7)$.

\end{itemize}

\subsection{Deterministic Function Categorization}\label{sec:method:rule}
This section defines five invocation types based on typical invocation patterns and categorizes functions accordingly. Table~\ref{tab:define} shows an overview of the five types.
The guiding principle for definition is from easy to difficult, also serving as the categorization priority: if a function fits a former type, it will not fit any latter type.
Our introduction follows this order.

\begin{table*}[htb]
    \small\centering
    \caption{A brief overview of the well-defined function types.}
    \vspace{-0.1in}
    \begin{tabular}{cccc}
    \toprule
    \textbf{Type name} & \textbf{Characteristics} & \textbf{Definition} & \textbf{Predictive values} \\
    \midrule
    \multirow{2}*{Always warm} & \multirow{2}*{Almost invoked all the time} & {Invoked at every time} & \multirow{2}*{--} \\
    && Inter-invocation time $\leq$ observing time\textperthousand & \\

    \midrule
    \multirow{2}*{Regular} & \multirow{2}*{Almost invoked periodically} & {(Processed) $P_{95}(\{WT\})-P_5(\{WT\}) \leq 1$} & \multirow{2}*{Median of WTs} \\
    && CV of $\{WT\} \leq 0.01$ & \\

    \midrule
    \multirow{2}*{Appro-regular} & \multirow{2}*{Invoked quasi-periodically} & (Processed) the count of Mode$_n(\{WT\}) \geq$  & \multirow{2}*{Mode$_n(\{WT\})$}\\
    && $0.9 \times$ sequence length & \\

    \midrule
    Dense & Frequently invoked &   $P_{90}(\{WT\}) \leq$ a small constant & $[\min, \max]$ of $\text{Mode}_k(\{WT\})$ \\

    \midrule
    \multirow{2}*{Successive} & \multirow{2}*{Successively invoked} & $\min(\{AT\}) \geq \gamma_1$; $\min(\{AN\}) \geq \gamma_2$ & \multirow{2}*{--} \\
    && $\gamma_1 < \gamma_2$ & \\
    
    \bottomrule
     \end{tabular}\label{tab:define}
     \vspace{-0.1in}
\end{table*}

% Structure:
% behavior summary
% a,b, c patterns --> observed predictable behavior
% Example if necessary
% Definition
% Predictive value/ provision strategy

\subsubsection{Always warm} 
This type describes consistently active functions, such as long-running operations and hyperfrequent calls, which usually involve durable or timer-triggered functions (whose interval is small enough).
For example, functions in continuous integration/deployment (CI/CD) pipelines are consistently invoked to automate code building, testing, and deployment.
A function is defined as ``always warm'' if 1) it is invoked at every sampling time or 2) the sum of inter-invocation time is $\leq$ one-thousandth the observing time.
Such functions are undoubtedly always loaded.

\subsubsection{Regular}
Regular task processing, usually using timer-triggered functions, is a basic functionality (e.g., polling).
Yet, their actual invocations may not be strictly periodic (with a constant WT) since there are undesired fluctuations:
1) The exact first/last WTs during an observing period are almost impossible to record.
2) Periodically generated events can be blocked or delayed by contingencies such as concurrency limits, network connectivity issues, etc. 
3) Other events can invoke a mostly regularly invoked function occasionally.

Here comes the slacked definition:
A ``regular'' function satisfies either 1) the difference between the 5th and 95th percentile of its WT sequence is $\leq 1$ or 2) the coefficient of variation of WTs is close to zero ($\leq 0.01$ in practice). 
Otherwise, we remove the first and last WTs and re-check if the function follows the definition.
If the answer is still no, we apply another slacking rule by merging adjacent small WTs.
In particular, for each WT closely valued to the WT mode, its adjacent small WTs are gradually merged until reaching \textcircled{{\scriptsize 1}} the sequence's end or \textcircled{{\scriptsize 2}} another WT close to the mode or \textcircled{{\scriptsize 3}} an already merged WT. 
In this way, a WT sequence valued by $(1439, 1438, 1, 1439, 1438, 1)$ is processed to be $(1439, 1439, 1439, 1439)$, and then it seems ``regular''.
With a processed WT sequence satisfying the above definition, the function belongs to the ``regular'' type.
We partially eliminate accidental factors by processing WTs with slacking rules and modeling essential invocation behaviors.
Finally, we record the median of WTs as the \textit{predictive value} for invocation prediction, leveraged in Section~\ref{sec:method:provision}.

\subsubsection{Approximatively regular} 
This is the derived type from ``regular".
A regular function may experience invocation variance and long-term disturbance.
For example, a data-processing function is expected to receive updates from a data station every three minutes (IoT Hub trigger functions). Yet, limited by the data transmission capability, the function is actually invoked every 3--5 minutes.
We define such functions as ``approximatively (appro-)regular''. 
Particularly, a function is ``appro-regular'' if the occurring count of the first $n$ frequently appearing values (modes) of WTs is $\geq$ a large percent (90\% practically) of the WT sequence's length. $n$ is a pre-defined integer.
The predictive values for these ``appro-regular'' functions are the first $n$ modes.

\subsubsection{Dense}
This characterizes irregular but frequent invocations with intermittent idle durations.
``Dense'' functions can involve several triggers: queues (or service bus using the queue flavor) for frequent messages in asynchronous processing, influenced by variations in message arrival times; Cosmon DB triggers for frequent real-time data processing; HTTP triggers for processing frequent HTTP requests.
Functions with a 90$^{th}$ percentile of the WTs $\leq$ a small constant are defined as ``dense.''
The range of the first $k$ modes derives predictive values, \textit{i.e.} $[\min(\text{Mode}_k(\text{WTs}))$, $\max(\text{Mode})_k(\text{WTs})]$, where $k$ is an empirical integer.
Such a function should be unloaded only if its idle time is larger than the constant.

\subsubsection{Successive}
This describes inactive functions experiencing consecutive invocations in a short timeframe (temporal locality) before returning to an inactive state.
It can involve HTTP trigger functions (e.g., bursty social media trends), storage trigger functions (e.g., uploading files to Blob Storage), etc.
In practice, feedback loops engaged functions, caching mechanism implementations, and load balancing-involved applications can also exhibit such behavior.
We define this as $\min(\{AT\}) \geq \gamma_1$ or $\{AN\} \geq \gamma_2$, where $\gamma_1 < \gamma_2$ and they are pre-defined lower bounds.
Predicting the start of such an invocation wave is highly difficult.
Nevertheless, we can tolerate the first invocation and keep the function alive until the wave ends. With limited cold starts, we save lots of memory.

\subsection{Indeterminate Function Assignment}\label{sec:method:difficult}

This section handles functions that do not satisfy the above five definitions using two strategies: \textcircled{{\scriptsize 1}} ``forgetting'' previous records to re-check whether the not-too-long history invocations satisfy the five definitions, and \textcircled{{\scriptsize 2}} assigning the remaining functions with three extra types.

\subsubsection{Forgetting}
As discussed in $\S$\ref{sec:motivation:challenges:evolution}, function invocations experience concept shifts. For example, a regular invoked function can be categorized as “always warm” if the interval is adjusted to be smaller than the recording unit.
Therefore, the near-the-present records should be given priority.
We slice out the sequential invocation observations in days, from the beginning to the half.
Then, we gradually remove the older observations from the original $d$-day observations and check whether invocations in 2nd--$d$th days conform to existing definitions. 
If not, we check the invocations in 3rd--$d$th days, and so on, until the $\lfloor d/2 \rfloor$th day.

\subsubsection{Assigning}
The remaining indeterminate functions are assigned to three types according to the validation results:

\begin{itemize}
    \item[D1.] \textit{Pulsed} describes functions showing less obvious temporal locality in invocations than ``successive" ones.
    Similarly, we allow a cold start at the first invocation time and keep the function warm until its ideal time reaches a pre-defined threshold.
    
    \item[D2.] \textit{Correlated} invocations are predicted based on other functions.
    Functions usually work in a logic workflow topology, interacting with other functions. This is present in several common application patterns, including function chaining (where a sequence of functions executes in a specific order), fan-out (where a function calls multiple functions in parallel), and fan-in (where a function waits for multiple functions to finish)~\cite{AzureScenario}.
    We define such functions using $T$-lagged co-occurrence rate ($T$-COR), derived from COR in $\S$\ref{sec:motivation:insight}. $T$-COR is the COR between the target invocation sequence and an $T$-time-lagged invocation sequence of its candidate functions (those sharing an application/user).
    The lagged COR can better reflect how the invocation of candidate functions indicates near-future invocations of the target function.
    If the $T$-COR reaches a threshold (practically 0.5, $T \leq 10$), the candidate invocation is an important predictive indicator, and these two functions are linked.
    A function can have multiple linkages with different functions, including categorized and uncategorized ones.

    \item[D3.] \textit{Possible} functions are infrequently invoked but have at least one mode of WTs (appears more than once). By taking the modes as predictive values, it is possible to predict invocations, though such prediction is deemed to be insufficiently satisfactory.

\end{itemize}

If any of the three strategies yields both the least cold starts and the minimum wasted memory during validation, the indeterminate function is directly assigned accordingly.
Otherwise, the rate of rise is decided. 
Denote the cold starts of the above three strategies as $(cs_1, cs_2, cs_3)$ and the corresponding wasted memory are ${wm_1, wm_2, wm_3}$.
Assume definition D1 delivers the minimum cold starts while strategy D2 delivers the minimum wasted memory.
We compute the rise rates: $\Delta cs=(cs_2-cs_1)/cs_1$ and $\Delta wm=(wm_1-wm_2)/wm_2$. 
If $\Delta cs \times \alpha \leq \Delta wm $, then the function is assigned to D1, \textit{i.e.} ``plused''; otherwise, D2 prevails, where $\alpha \in (0,1)$ is a scaling factor, and the smaller $\alpha \in (0,1)$, the more importance is put on cold starts than wasted memory.

Most functions have been categorized so far, except those that have never been invoked in the validation. 
We simply leave them as ``unknown'' since they are likely very infrequently invoked and hardly have meaningful predictive indicators. 
Some unknown functions may be categorized during the online provision based on our adaptive strategies illustrated in the next section ($\S$\ref{sec:method:adaptive}).

\subsection{Adaptive Strategy Application}\label{sec:method:adaptive}
This section designs two adaptive strategies to handle concept shifts during the online provision: one adjusts the predictive values; the other is based on inter-function correlation. 
These strategies can further address some unknown or unseen functions (those that never appear in the training data) with meaningful patterns during the online provision.
The application of these strategies is also presented in Figure~\ref{fig:overview}, where the invocation prediction and the prediction values are adaptively updated upon online invocations.

\subsubsection{Adjusting}
This adaptive strategy contains three steps:
\begin{itemize}

    \item[S1.] Record the WTs during the online provision. If there are enough WTs, then initiate the adaptive updating.
    
    \item[S2.] Update the predictive values if the corresponding values computed from newly collected WTs change significantly with the mean of the old and new ones.
    
    \item[S3.] If the new WTs for an unknown or unseen function conform to previous definitions with enough samples, categorize the function as the corresponding type.
    
\end{itemize}
S2 requires more explanations.
We record the predictive values for four types: `regular'', ``appro-regular'', ``dense'', and ``possible''.
Their predictive values are the median, the first $n$ modes, the range of the first $k$ modes, and the WT values occurring more than twice, respectively.
Take ``regular'' as an example. 
Suppose the absolute difference between the median of offline WTs (\textit{i.e.} the predictive value) and that of online WTs is larger than the standard of offline WTs. The new predictive value is updated by the mean of the old median and the new median.
Other types adopt a similar adjusting strategy. 

\subsubsection{Online correlation}
This strategy correlates unseen functions with known functions or appeared unseen functions, as the ``correlated'' strategy does.
To speed up the computation and obtain informative indicators, we only consider candidate functions sharing the same trigger with the target function (\textit{aka.} unseen function).
Initially, if one of the candidates is invoked, we also pre-load the target function.
Afterward, we gradually remove less correlated candidates.
Again, we adopt COR to measure the degree of correlation.
We count the pair-wise target-candidate COR at each slot and record the maximum.
If the difference between a COR and the maximum is large enough, the corresponding candidate is kicked out from consideration unless its COR returns close to the maximum.

\subsection{Next Invocation Prediction and Provision}\label{sec:method:provision}
This section predicts invocations and pre-loads functions based on the predictive values or indicators.
``(Appro-)regular'' functions have discrete predictive values and ``dense'' functions use continuous ones.
In terms of ``possible'' functions, if the range of predictive values is larger than a threshold, the values are regarded as discrete; otherwise, we consider continuous integers inside the range of predictive values.
The predicted invocation times are naturally the last invoked time added by each of the predictive values.

For the online provision, if one of the predicted invocation times falls in $[t-\theta_{prewarm}, t+\theta_{prewarm}]$ with a pre-defined parameter $\theta_{prewarm}$ at the time of $t$, the function will be pre-loaded. 
If a loaded function's current WT is $\geq \theta_{givenup}$, the function will be evicted from the memory, where the parameter $\theta_{givenup}$ differs among function types.
Algorithm~\ref{algo:provision} shows our provision optimization, which is rule-based with good scalability and implements our defined strategies faithfully.
Every time the algorithm returns the updated MemSet, based on which we load function instances.

\begin{algorithm}[htbp]
    \small
    \caption{The provision algorithm of \tool.} \label{algo:provision}
    \LinesNumbered
    \KwIn{FList: hash ids for all functions; 
        FState: recording the necessary information about all functions, such as the current WT, history WTs, predictive values, the time of the last invocation, etc; 
        Invo$^{(t)}$: invocation numbers of each function at the time of $t$; 
        MemSet: hash ids of loaded functions; 
        UCorr: the correlations of unseen functions.
    }
    \KwOut{The updated MemSet, FState and UCorr}

    \SetKwFunction{FDeg}{Provision}
    \SetKwProg{Fn}{Function}{:}{End}
    \Fn{\FDeg{FList, FState, Invo$^{(t)}$, MemSet, UCorr}}{
    \For{$f \in$ FList}{
        \eIf{Invo$^{(t)}$[$f$] $>$ 0}
            {
                FState[$f$].last\_invoked $\leftarrow t$; \\ 
                FState[$f$].update\_WTs(FState[$f$].current\_WT); \\
                FState[$f$].current\_WT $\leftarrow 0$; \\ 
                
                // Adaptively adjusting predictive values.\\
                FState[$f$].update\_predictive\_values($t$); \\
                
                \If{$f \notin$ MemSet}{
                    FState[$f$].cold\_start\_record($t$); \\
                    MemSet.add($f$); \\}
            }{ %else
                FState[$f$].current\_WT += 1;\\
                pre\_load\_flag $\leftarrow$ FState[$f$].pre\_load(FState[$f$].$\theta_{prewarm}$); \\
                
                \uIf{pre\_load\_flag is \textbf{False} \textbf{AND} FState[$f$].current\_WT $\geq$ FState[$f$].$\theta_{givenup}$}{ 
                    MemSet.remove($f$); \\
                    }
                \uElseIf {pre\_load\_flag is \textbf{True}}{
                    MemSet.add($f$); // Pre-loading. \\
                }
                
            }
        // Adaptively processing unseen functions. \\
        UCorr.update(); 
    }
    \Return{MemSet, FState, UCorr}
    }
\end{algorithm}

\section{EVALUATION}\label{sec:exp}
We evaluate \tool by answering four research questions:
\begin{itemize}[leftmargin=12pt, topsep=0pt]
\item \textbf{RQ1}: How effectively does \tool decrease cold starts?
\item \textbf{RQ2}: How much memory waste and computation overhead does \tool incur?
\item \textbf{RQ3}: How does \tool trade off memory waste with latency reduction?
\item \textbf{RQ4}: How do complementary designs influence \tool?
\end{itemize}

\subsection{Experiment Settings}\label{sec:exp:setting}
We evaluate \tool on the most widely used industrial dataset~\cite{AzureDataset} released by Microsoft Azure Function~\cite{ServerlessWild} with real-world invocation traces. 
The dataset contains the invocation counts per minute for 14 days. The first 12 days are used for pattern modeling (training), and the last two days are used for the simulation, conducted on a workstation with an 8-core Intel i5-3470S CPU and 16 GB memory.
The records involve 15,097 users, 24,964 applications, and 83,137 functions. 71,616 functions appear in the training data, 39,388 functions appear in the simulation data, and 743 never appeared during training.

We adopt the simulation principles following~\cite{ServerlessWild}.
First, assume all executions finish within one minute since 1) most (96\%) functions have very short execution time (less than 60 seconds), and 2) we can thereby calculate the worst-case wasted resource time.
Second, we assume that cold-start latency for each function is the same, so we only need to care about the number of cold starts and the wasted memory time.

\subsubsection{Baselines}\label{sec:exp:baselines}
We compare \tool with five state-of-the-art baselines applied to the application layer. 
Approaches involving system renovations are out of this paper's scope and will be discussed in Section~\ref{sec:review}.
Our baselines include FaaSCache~\cite{FaasCache}, Defuse~\cite{Defuse}, Hybrid~\cite{ServerlessWild}, and a fixed keep-alive policy.
We reproduce these methods as they do not provide open-source code by strictly following the original papers and referring to a reproduction attempt~\cite{Reproduce}.
Note that the original Hybrid method works at the application unit (Hybrid-Application, HA), so we derive a Hybrid-Function method by employing its design on the function granularity (Hybrid-Function, HF), following~\cite{Defuse}.
All the parameters are set according to their original papers. The fixed keep-alive policy adopts a length of 10 mins. FaaSCache requires a pre-defined memory limit, so we adopt the maximum memory size of \tool during the whole simulation.

\subsubsection{Metrics and parameters}\label{sec:exp:metrics}
To quality the cold-start optimization, we measure the function-wise (application-wise for HA) \textit{cold-start rate (CSR)}, \textit{i.e.} the number of cold starts divided by the number of invocations. 
We apply wasted memory time (\textit{WMT}, $\S$\ref{sec:background:challenge}) to gauge the idle resource waste.
Naturally, the lower the CSR or WMT, the fewer cold starts or wasted resources, the better.
We also monitor the \textit{effective memory consumption ratio (EMCR), which measures the fraction of invoked function instances relative to the total loaded instances on each host machine, serving to assess resource efficiency. A higher EMCR signifies wiser memory allocation, as it indicates a greater proportion of memory is used by active instances rather than remaining idle.}
As for the parameters, we set $\theta_{prewarm}$ as two. 
The $\theta_{givenup}$ for ``dense'' and ``plused'' is five, whereas one for the other types. 
Section~\ref{sec:exp:trade-off} further discusses the impact of these pre-defined parameters.

\subsection{RQ1: Effectiveness in Cold-Start Reduction}\label{sec:exp:cold-start}

\begin{figure}[htb]
    \centering
    \vspace{-0.13in}
        {\includegraphics[width=0.98\linewidth]{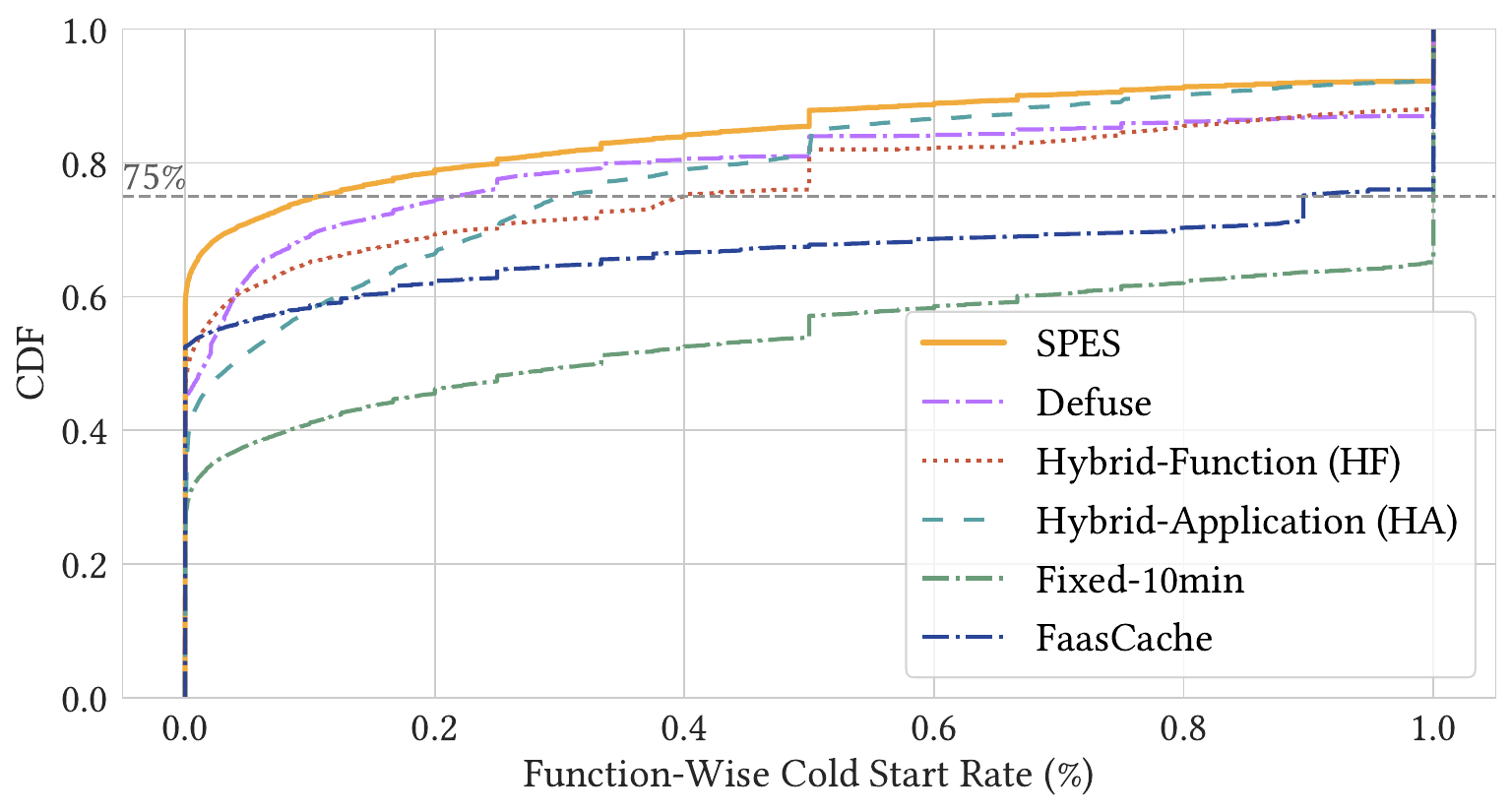}}
        \vspace{-0.1in}
    \caption{Cold start behaviors of \tool and its competitors.}
    \vspace{-0.1in}
    \label{fig:CDF_cold_start}
\end{figure}

Figrue~\ref{fig:CDF_cold_start} displays the cumulative distribution (CDF) of CSR under the provision decisions of \tool and baselines. 
With a fixed y-axis value, the line representing \tool is positioned to the left of other baseline lines, which indicates that \tool consistently leads to fewer cold starts across a variety of functions, each corresponding to different invocation frequencies.
In particular, \tool reduces the 75th percentile cold-start rate (Q3-CSR, for simplicity) from 0.215 to 0.108 compared to Defuse, the best-performing baseline, achieving 49.77\% improvement, and reduces 75-CSR by 64.06\%--89.20\% compared to other baselines. 
We care more about Q3-CSR because infrequently invoked functions benefit most from optimization~\cite{ServerlessWild}.
Moreover, 57.99\% functions experience no cold starts with \tool, indicating that \tool can allow most functions to be warmly invoked. In contrast, 25.61\%--52.59\% functions completely experience warm startup using baselines where FaaSCache is the best one.
On the other hand, regarding infrequently invoked functions, \tool reduces the 90th percentile CSR by 19.87\% compared to the best-performing baseline, HA. 
Such improvements imply that \tool exhibits significant optimization performance for both frequently and infrequently invoked functions, while no baseline can achieve second-best performance with different percentile CSRs.

\begin{figure}[htbp]
\centering
\subfigure[Normalized memory usage]{
\begin{minipage}{0.22\textwidth}
\centering 
\includegraphics[width=0.95\linewidth]{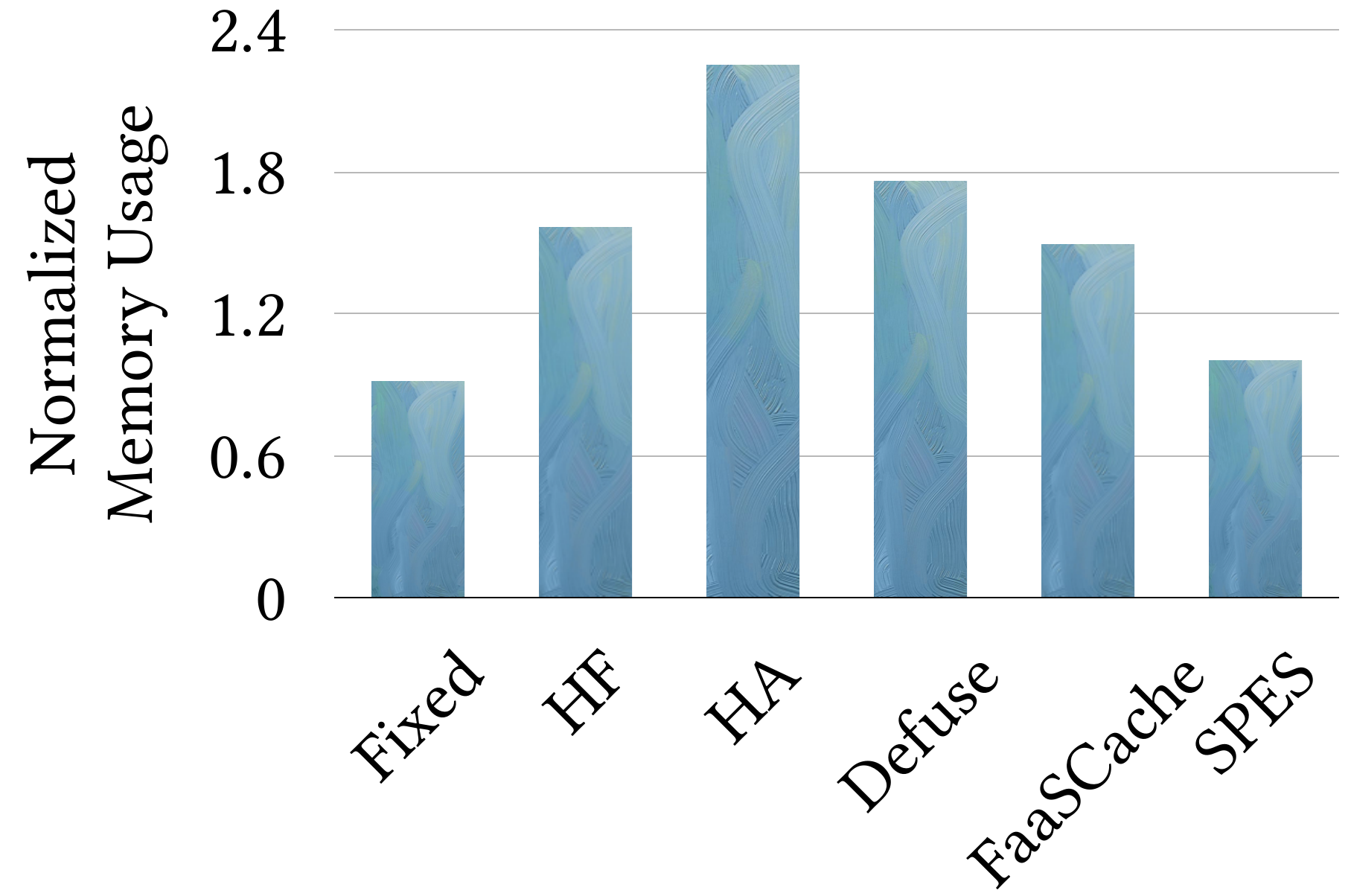}
\label{fig:normalized_mem_usage}
\end{minipage}
}
\subfigure[Percentage of always-cold functions]{
\begin{minipage}{0.22\textwidth}
\centering 
\includegraphics[width=0.95\linewidth]{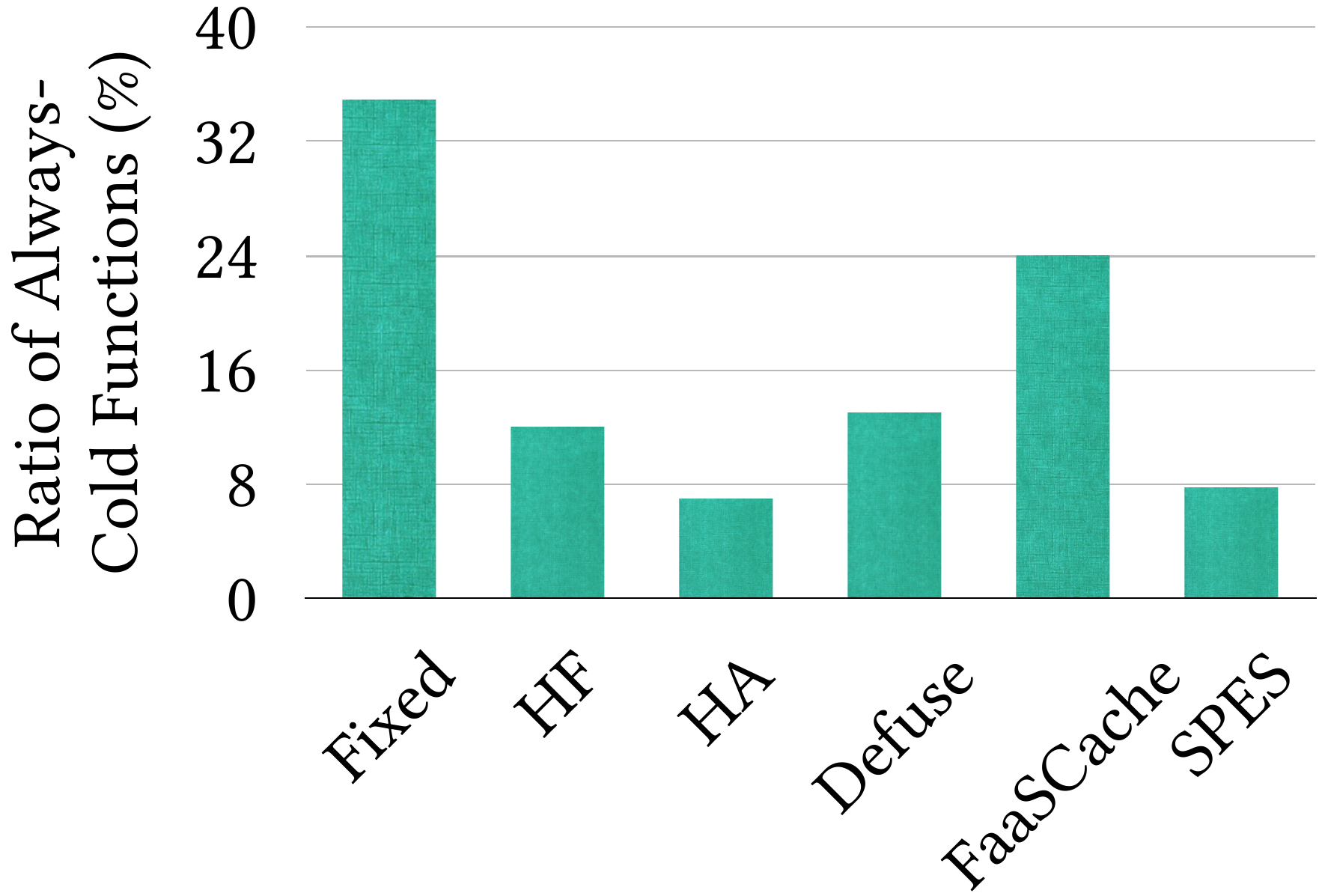}
\label{fig:total_cold_perc}
\end{minipage}
}
\vspace{-0.1in}
\caption{\tool is memory-efficient compared to most baselines with relatively few always-cold functions.}
\vspace{-0.1in}
\label{fig:high_csr_normalized_memory}
\end{figure}

Furthermore, \tool significantly reduces ``always-cold'' functions, which always experience cold starts upon invocations (with CSR equal to 1.0).
Figure~\ref{fig:total_cold_perc} further presents the always-cold percentage, and that of \tool is just less than 8\%.
Among the baselines, HA has the fewest always-cold functions, which is closest to that of \tool, whereas Defuse and HF, both function-grained methods, increase always-cold functions dramatically.
This can be attributed to infrequently invoked functions.
About 3.82\% functions are invoked less than twice during training, and 6.14\% are only once during the simulation, so always-cold functions seem inevitable with limited records.
HA mitigates this issue by grouping and loading functions together, yet resulting in more memory consumption and waste.
\tool instead connects unseen and unpredictable functions with known functions using trigger, application, and user information and gradually updates the associated functions based on actual invocations. 
Hence, without much extra memory, \tool performs close to HA.

In addition to cold start reduction, Figure~\ref{fig:normalized_mem_usage} presents the memory usage normalized by the averaged one of \tool.
\tool's memory usage is only 8.08\% more than the most resource-efficient method, the fixed keep-alive policy, and saves 36.07\%--55.55\% memory compared to the other baselines on average. 
It only consumes about half of the memory of Defuse, the best baseline for cold start reduction.

\begin{figure}[htb]
    \centering
    \vspace{-0.1in}
        {\includegraphics[width=0.95\linewidth]{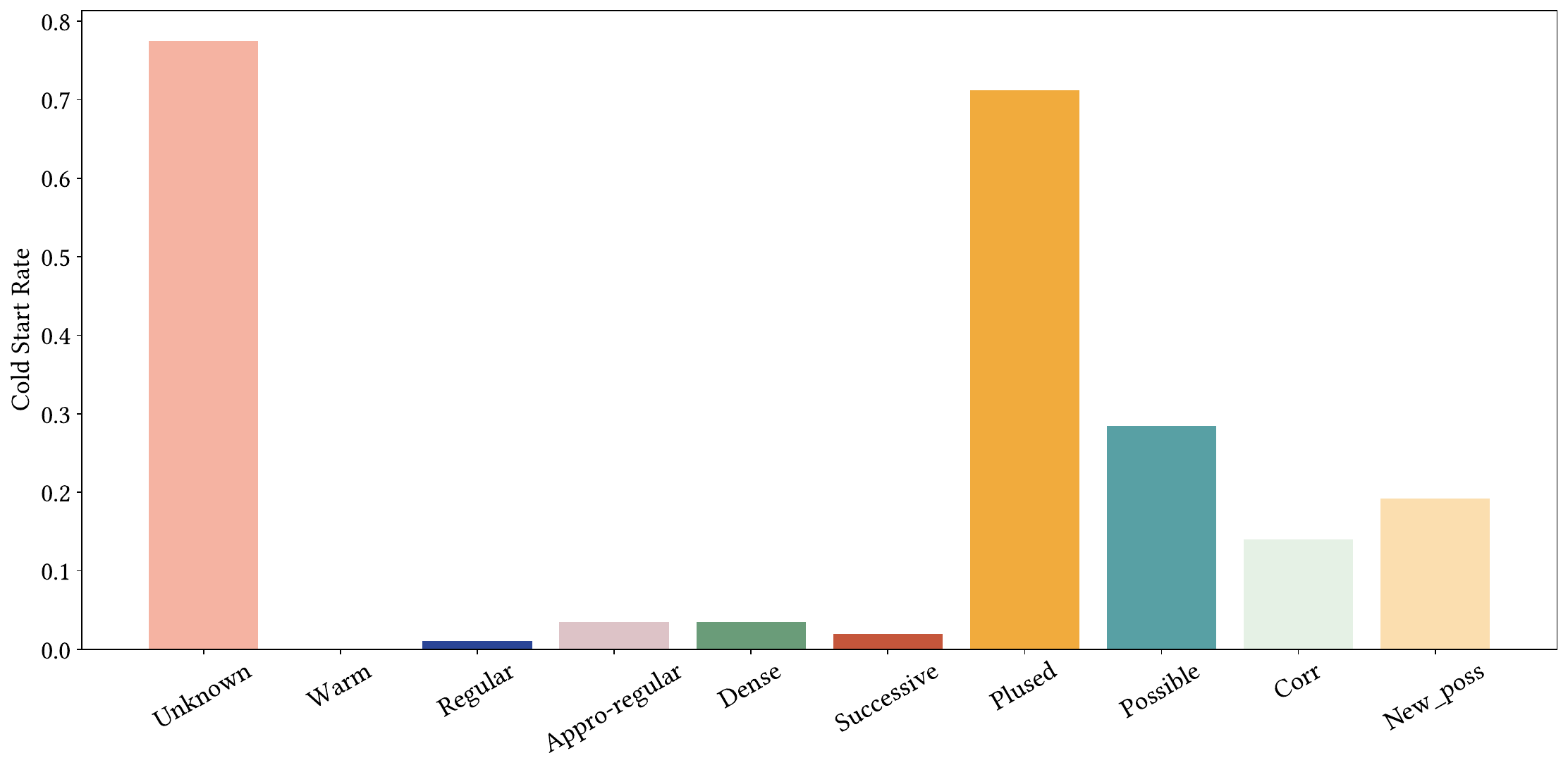}}
        \vspace{-0.1in}
    \caption{Averaged cold start rate of each type.}
    \vspace{-0.1in}
    \label{fig:cold_start_per_type}
\end{figure}

Figure~\ref{fig:cold_start_per_type} shows the averaged CSR among different types. As the simulation period is not long enough, we only categorize unknown functions into the type of ``possible'', denoted by ``newly-possible''. 
%Other types require more samples to rule out arbitrary factors.
We can see that ``unknown'' functions contribute the most to cold starts, and ``pulsed'' functions also incur high CSRs. This is attributed to insufficient historical invocation records.
Actually, \tool intentionally connives a cold start when an ``unknown'' or ``pulsed'' invocation arrives after a long idle time.
Though we can leverage less predictive information, such as the averaged WT, to predict the next invocation or even keep such functions always warm, this can lead to considerable unbillable and wasted memory, undesired for FaaS providers.
This outcome is deliberating on the trade-off between performance and resource allocation.
Such an issue can be mitigated with more invocation histories revealing predictive indicators for better provision.

Meanwhile, though ``correlated'' and ``possible/newly-possible'' functions also lack predictive information with infrequent invocations, their invocations are more likely to be predicted. Thus, we connect some of them with deterministic functions or adaptively extract meaningful behavior indicators from incoming invocations. 
In this way, their cold starts are suppressed effectively. We will further discuss the usefulness of such strategy designs in Section~\ref{sec:exp:ablation}.

\subsection{RQ2: Wasted Memory Time and Overhead}\label{sec:exp:wasted-memory-time}

\subsubsection{Wasted Memory Time}
\begin{figure}[htbp]
\centering
\subfigure[Normalized wasted memory time (WMT, lower is better)]{
\begin{minipage}{0.22\textwidth}
\centering 
\includegraphics[width=0.95\linewidth]{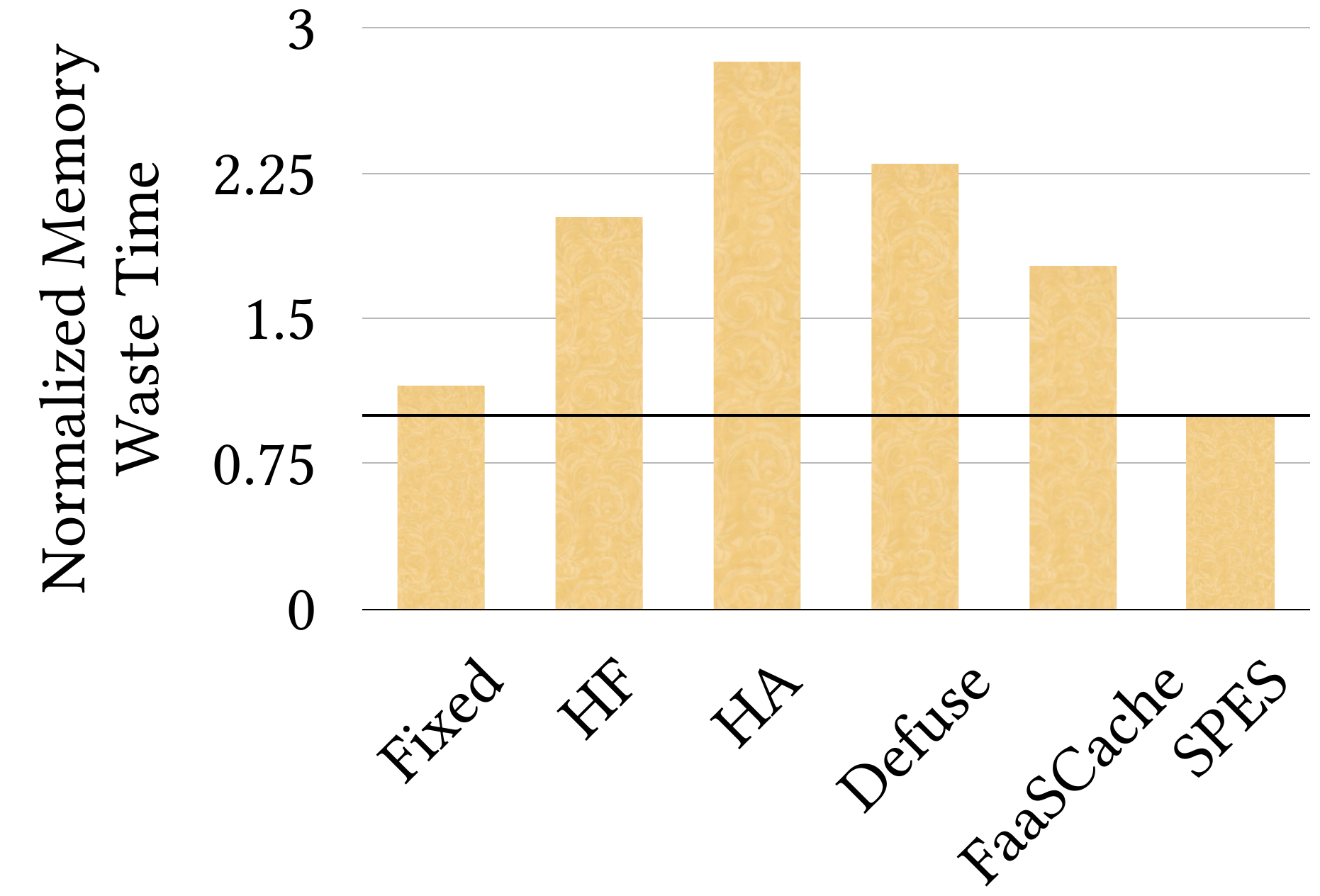}
\label{fig:waste_memory_time}
\end{minipage}
}
\subfigure[Effective memory consumption ratio (EMCR, higher is better)]{
\begin{minipage}{0.22\textwidth}
\centering 
\includegraphics[width=0.95\linewidth]{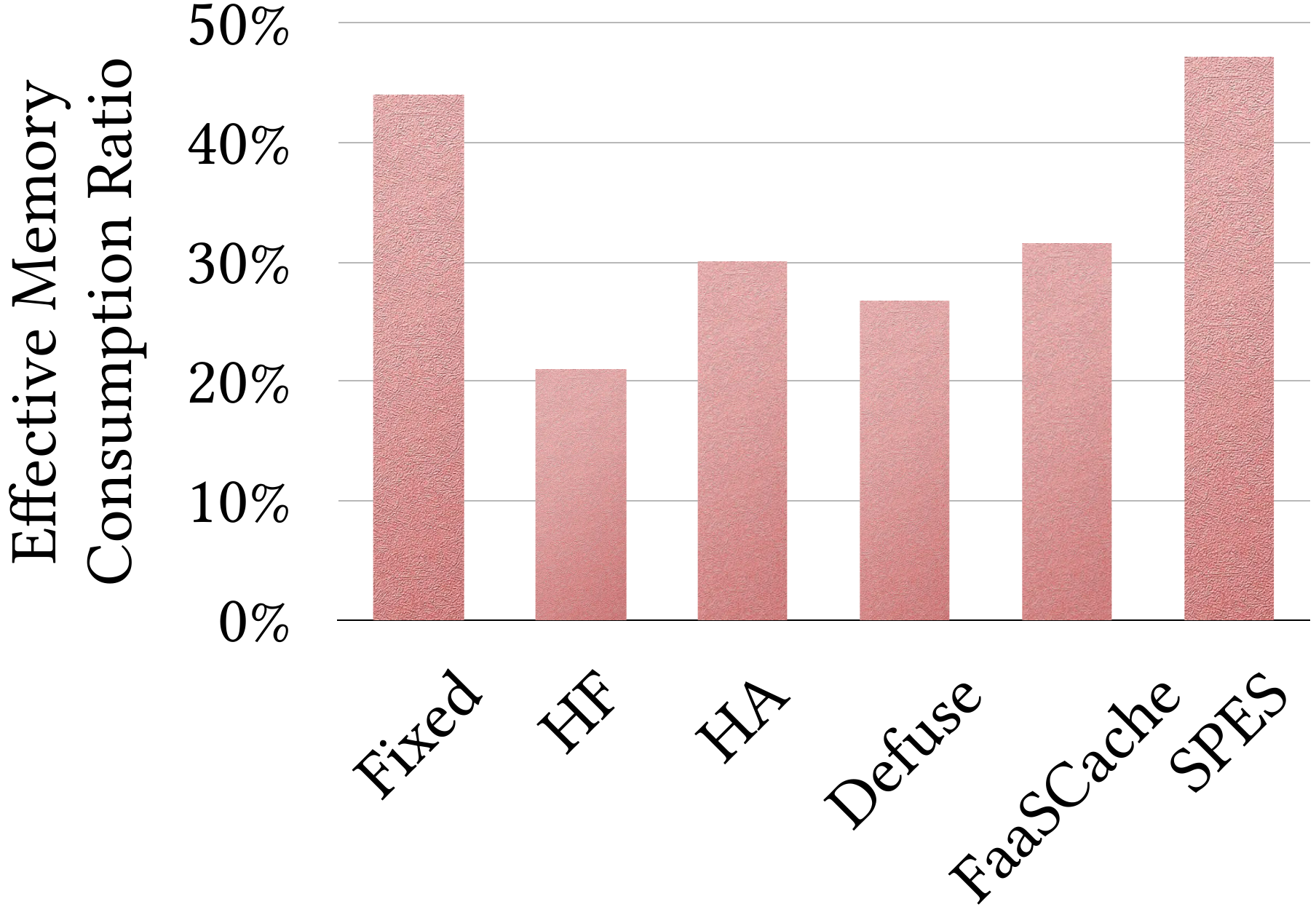}
\label{fig:memory_usage_ratio}
\end{minipage}
}
\vspace{-0.1in}
\caption{\tool significantly reduces memory waste and achieves more effective memory consumption.}
\vspace{-0.1in}
\label{fig:wasted_memory_and_usage}
\end{figure}

Figure~\ref{fig:waste_memory_time} presents that \tool significantly decreases the wasted memory time by 10.89\%--63.50\% compared to all baselines. Particularly, compared to Defuse, the most effective baseline on cold-start reduction, we reduce 57.06\% of the WMT.
Figure~\ref{fig:memory_usage_ratio} also demonstrates that \tool efficiently uses memory resources, whose EMCR is 46.32\%, 5.20\%--120.89\% higher than compared approaches.
The success of \tool can be attributed to careful pattern modeling and differentiated strategies.

\begin{figure}[htb] 
    \centering
    \vspace{-0.1in}
        {\includegraphics[width=0.98\linewidth]{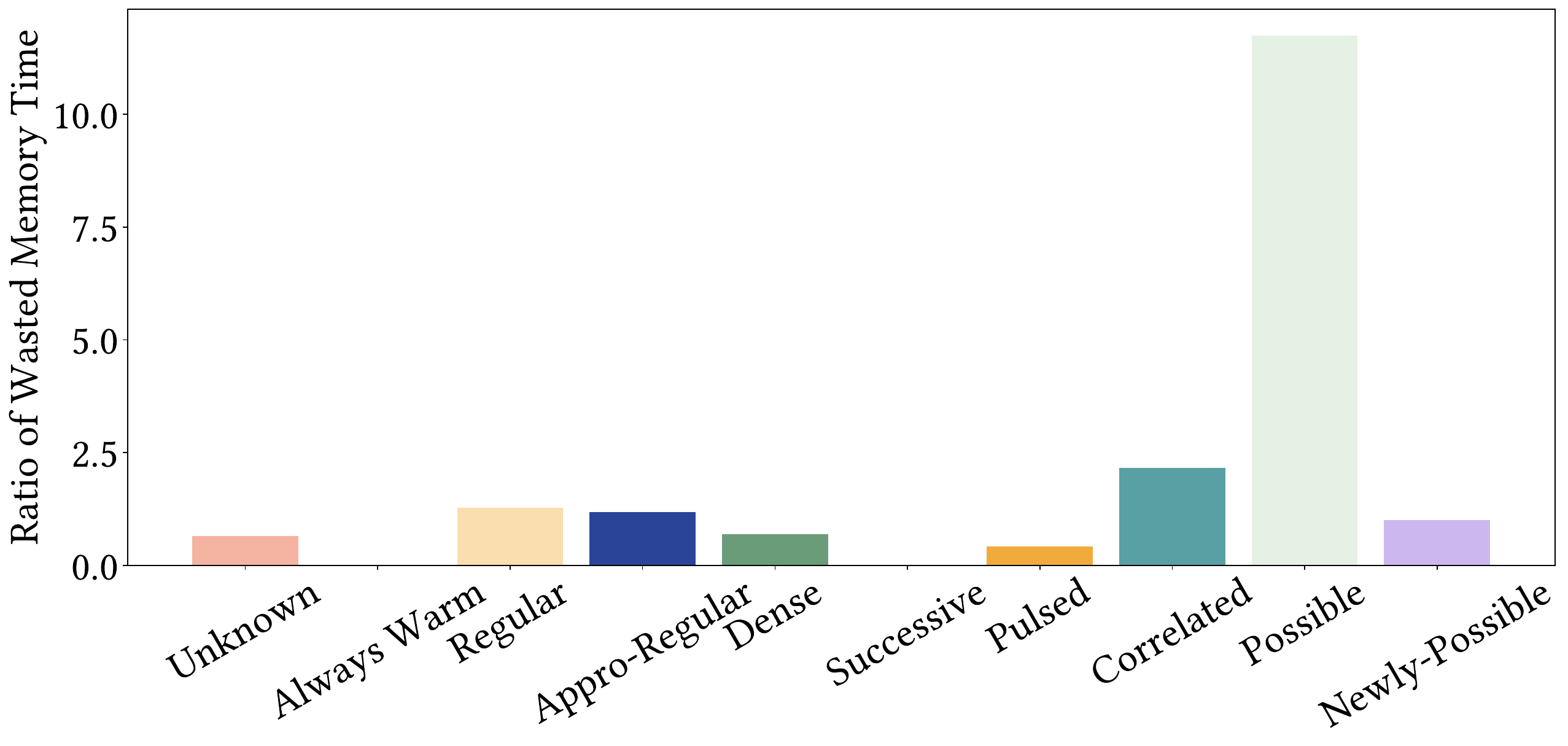}}
        \vspace{-0.1in}
    \caption{The ratio of WMT of each function type.}
    \vspace{-0.1in}
    \label{fig:ratio_WMT_per_type}
\end{figure}

To further demonstrate how each type contributes to WMT, we derive a metric called the \textit{ratio of WMT}, which is the WMT divided by the number of invoked times for each serverless function.
If a type contains more functions or function invocations, the bespoke scheduling strategy tends to make more proactive provisions, likely resulting in more WMT.
Thus, the ratio of WMT can better portray the accuracy of invocation prediction and the usefulness of pre-loading.
Figure~\ref{fig:ratio_WMT_per_type} shows the distribution of the ratio of WMT among different function types, from which we can see the ``possible'' functions have the highest probability of generating WMT. 
The predominant reason for this is the infrequent invocation of possible functions, which leads to a scarcity of patterns in the historical database and complicates establishing associations with other functions. Despite these challenges, we persist in our efforts to forecast and pre-warm the invocation of these functions, aiming to reduce cold starts as much as possible. 
Different from ``pulsed'' or ``unknown'' functions that are allowed to generate cold starts, we encourage aggressive prediction attempts for ``possible'' functions since the latter have at least a duplicated WT, enabling potential predictive value obtaining.
However, this strategy unavoidably results in augmented resource wastage.

\subsubsection{\tool Overhead}
Herein, we measure the additional latency induced by our implementation. The time complexity within each time window of our approach and the fixed keep-alive policy is $\mathcal{O}(n)$, where $n$ represents the number of functions, as it only requires accessing the corresponding function's category and its predictive value(s). For each invocation, the time complexity of pre-load/unload operations is merely $\mathcal{O}(1)$. Within each time window, the time complexity for the fixed keep-alive method is $\mathcal{O}(m)$, where $m$ is the number of loaded functions. In contrast, FaaSCache has a time complexity of $\mathcal{O}(\log m)$, which involves identifying the container with the lowest priority. Other methods, such as HA, HF, and Defuse, have higher complexities, mainly due to the computational bottleneck in updating histograms. 
The simulation results align with these expectations. 
Compared to the fastest fixed keep-alive baseline, which has an average overhead of 0.024 sec per minute, our method adds 0.44 sec's overhead per minute, mainly due to our more complex strategies for each provision action.
Contrasted with the second fastest, FaaSCache, our overhead is reduced by 6.8\%. 
In summary, our overhead is inconsequential compared to the typical latency found in most existing serverless platforms.

\subsection{RQ3: Trading-off resources and latency}\label{sec:exp:trade-off}

We control the trade-off through two parameters: $\theta_{prewarm}$ and $\theta_{givenup}$. 
As introduced in Section~\ref{sec:method:provision}, the former decides how long to pre-load a function with a predicted nearby invocation, and the latter decides how long to keep an idle function warm.
Intuitively, the larger these two parameters, the more likely a function is to pre-load or keep loaded with more memory usage and potential wasted memory, and the less cold starts.
In RQ1 ($\S$\ref{sec:exp:cold-start}), we set $\theta_{prewarm}$ as two, and the $\theta_{givenup}$ for ``dense'' and ``plused'' is five, whereas one for the other types. Our original simulation setting is denoted by the red star (\textcolor{luoshen}{$\star$}) in the following figures.
 
Figure~\ref{fig:prewarm} shows the trade-off under different $\theta_{prewarm}$, where a point $(x, y)$ represents using $x$-unit memory and obtaining the 75-CSR of $y$, under a certain $\theta_{prewarm}$.
The memory is normalized to that under the original setting.
The normalized memory usage and the 75-CSR are nearly linearly correlated.
We can conveniently choose a proper setting by controlling $\theta_{prewarm}$. 
Since the red star is below and the most distant from the fitting line, $\theta_{prewarm}=2$ is the optimal value.

As $\theta_{givenup}$ should be integers and different from each type, we simply multiply the original $\theta_{givenup}$ setting by 2, 3, 4, and 5, respectively.
The results are shown in Figure~\ref{fig:givenup}, where the linear relationship still approximately holds, but larger $\theta_{givenup}$s have less impact on cold start reduction. 
This indicates that keeping invoked functions too long is sub-optimal and idle functions should be evicted promptly.

\begin{figure}[htbp]
\centering
\subfigure[Parameter $\theta_{prewarm}$]{
\begin{minipage}{0.4\textwidth}
\centering 
\includegraphics[width=0.98\linewidth]{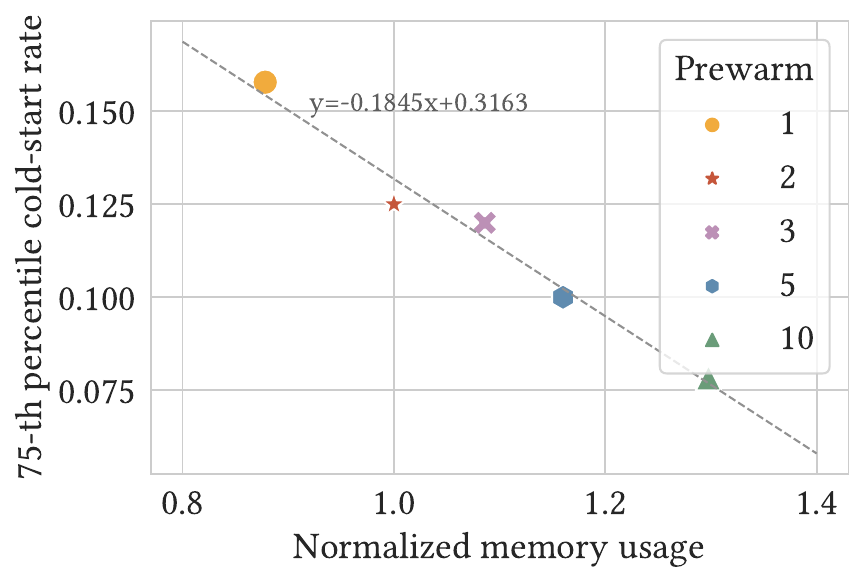}
\label{fig:prewarm}
\end{minipage}
}
\subfigure[Parameter $\theta_{given}$]{
\begin{minipage}{0.4\textwidth}
\centering 
\includegraphics[width=0.98\linewidth]{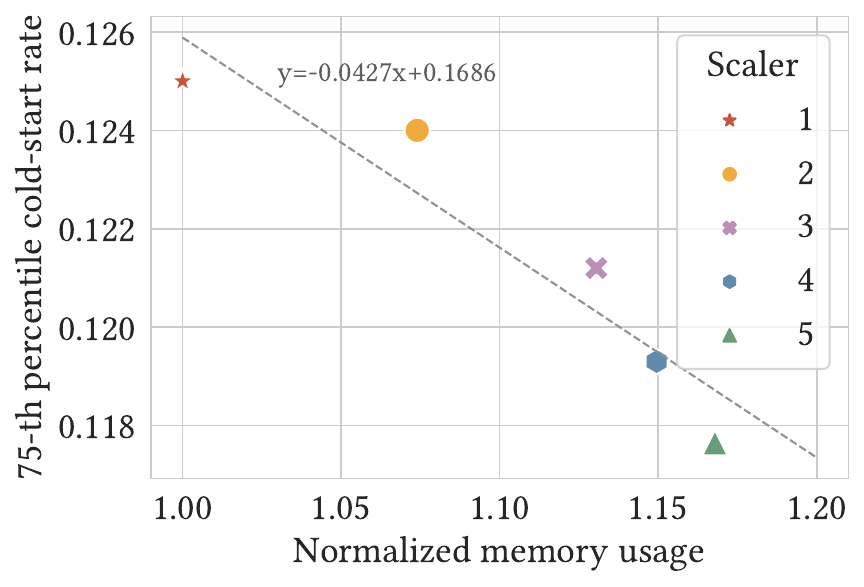}
\label{fig:givenup}
\end{minipage}
}
\vspace{-0.1in}
\caption{Under different $\theta_{prewarm}$, the memory usage and Q3-CSR exhibit an approximately linear relationship. Similarly, under different $\theta_{given}$, the linear relationship still approximately holds, yet dramatically increasing the memory does not guarantee cold start mitigation.}
\vspace{-0.1in}
\label{fig:prewarm-givenup}
\end{figure}

\subsection{RQ4: Impact of Strategy Designs}\label{sec:exp:ablation}
\subsubsection{Impact of inter-function correlations}\label{sec:exp:ablation:corr} 
\begin{figure}[htb]
    \centering
        \vspace{-0.08in}
        {\includegraphics[width=0.98\linewidth]{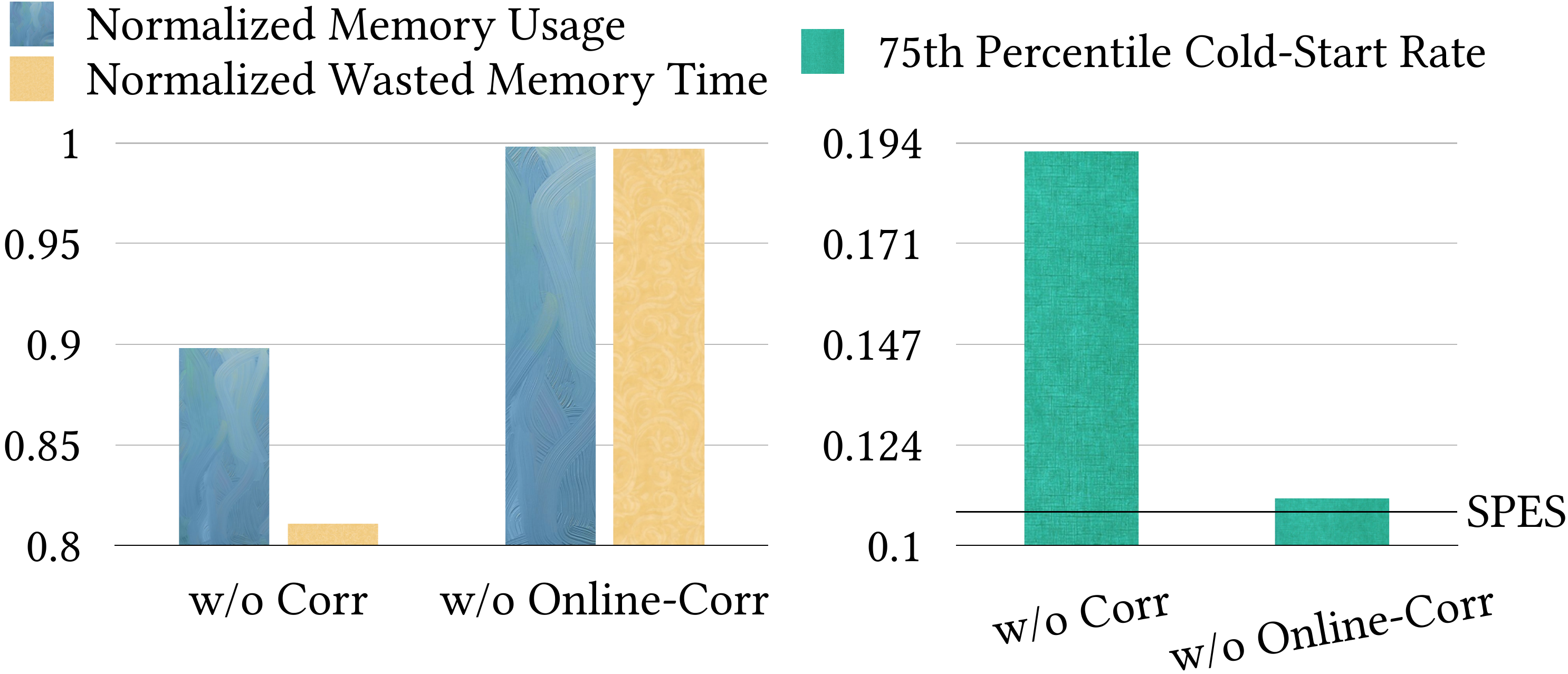}}
        \vspace{-0.1in}
    \caption{The correlation strategy contributes to the cold-start and memory-waste reduction.}
    \vspace{-0.12in}
    \label{fig:impact_corr}
\end{figure}

This section gauges our design in processing ill-informed functions by developing inter-function correlations.
As introduced in $\S$\ref{sec:method:difficult}, we propose a simple yet effective $T$\textit{-lagged co-occurrence rate} metric to connect ill-informed functions with categorized ones. Those with closely related known functions are ``correlated''.
This strategy is applied during both training and simulation.
Figure~\ref{fig:impact_corr} presents the impact of this strategy.
\textit{w/o Corr} means re-categorizing ``correlated'' functions into ``pulsed'', ``possible'' or ``unknown'' during training, but we still deal with unseen functions during the simulation.
\textit{w/o Online-Corr} denotes removing the simulation-applied strategy, \textit{i.e.} regarding unseen functions as ``unknown'' but still retaining the ``correlated'' functions obtained during training.
It is shown that the latter strategy slightly reduces the Q3-CSR, whereas the former makes a significant contribution.
We attribute the results to the influenced function number. 
4.71\% of the functions belong to ``correlated'' whereas only 1.89\% are unseen.

\subsubsection{Impact of designs regarding concept shifts}\label{sec:exp:ablation:shift}

\begin{figure}[htb]

    \centering
    \vspace{-0.1in}
        {\includegraphics[width=0.98\linewidth]{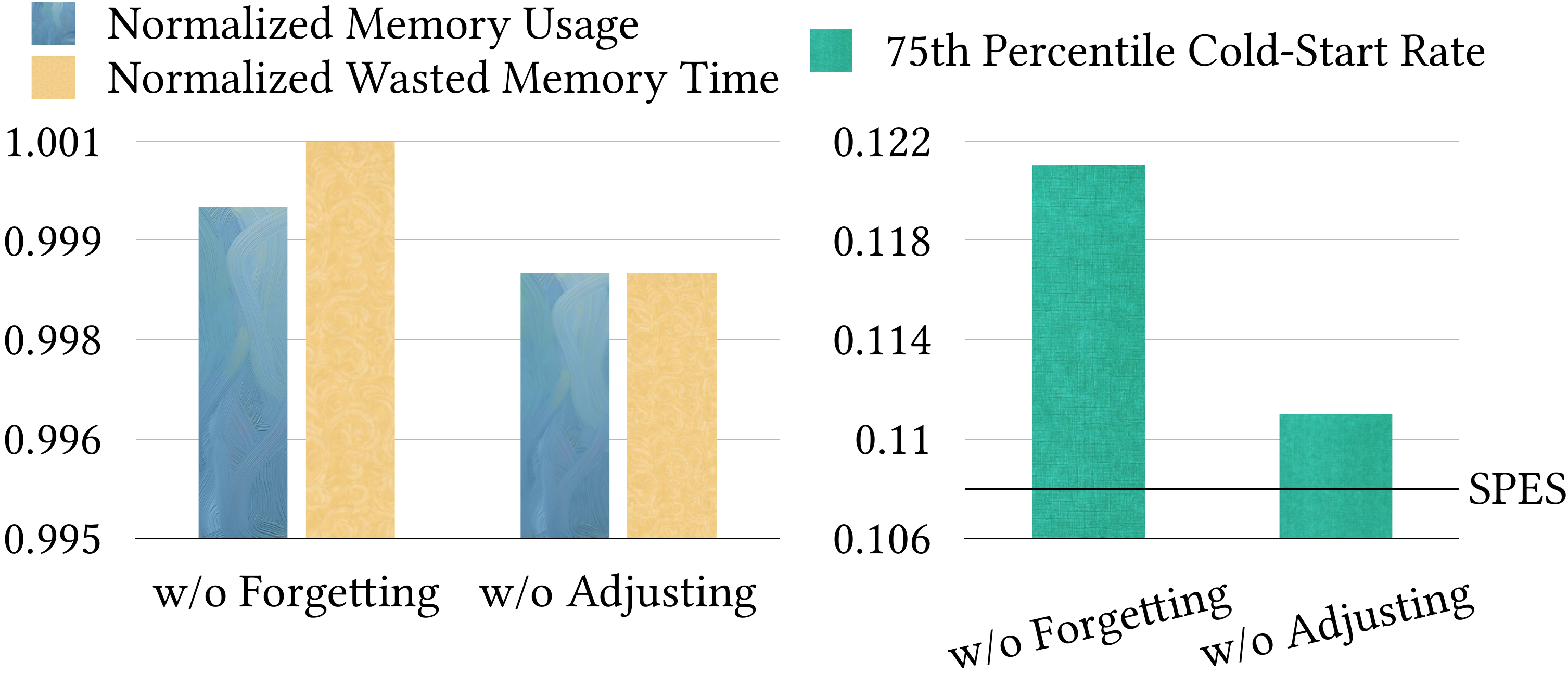}}
        \vspace{-0.1in}
    \caption{\tool' adaptivity benefits cold-start reduction.}
    \vspace{-0.1in}
    \label{fig:impact_shift}
\end{figure}

This section studies the impact of two designs regarding concept shifts:
1) \textit{forgetting}: forcing unknown functions to a defined type by ignoring older invocations while focusing more on recent data ($\S$\ref{sec:method:difficult});
2) \textit{adjusting}: adaptively adjusting the predictive values during the simulation ($\S$\ref{sec:method:adaptive}).
Figure~\ref{fig:impact_shift} depicts the effectiveness after omitting these two adaptive designs, respectively. 
Removing the second shifting strategy has a slighter impact. 
Similarly, this is attributed to the fact that the forgetting strategy involves more function categorization efforts.
We do not pre-load unknown functions, so the forgetting strategy categorizing 340 unknown functions has a larger impact. In contrast, the adjusting strategy only categorizes 174 unknown functions into the ``newly-possible'' type and updates the predictive values of 499 ``(appro-)regular/dense'' functions, resulting in less impact.
Nevertheless, both designs contribute to the effectiveness of \tool. These designs will play a greater value with more data and a longer simulation period (usually indicating larger shifts).

\section{Discussion}\label{sec:discuss}
\subsection{Limitations}
\subsubsection{Idealized Simulation Principles}
Our simulation adopts two key assumptions to simplify the process and mitigate dataset limitations. 
Firstly, we assume each function execution completes within one minute. For functions with longer execution, adjusting the counting unit as necessary to accommodate the vast majority (e.g., 95\%) of functions within this timeframe can ensure the assumption's relevance. 
For functions exceeding this one timeframe without new invocations—e.g., for a two-minute execution without activity in the second minute—we maintain memory allocation and increment the WMT despite no memory being actually wasted. 
Such deviation can particularly underestimate resource efficiency, especially for methods with frequent evictions, including ours and HF. 
Thus, we posit that such marginal deviations do not affect the overall integrity of our experimental findings.
Secondly, we assume all functions share the same cold-start latency. Our focus on function-specific metrics means that variations in cold-start latency across functions do not significantly affect our primary research insights, though a more detailed dataset could enhance our analysis of the aggregate cold-start latency and offer a deeper comparison of methodologies.

\subsubsection{Over-simplicity of memory consumption}
We simply assume that one node can accommodate all loaded function instances with infinite resources following~\cite{Defuse, ServerlessWild}, so we do not consider problems like worker communication, sandbox optimization, and load balancing. 
Otherwise, we may need to renovate the underlying system intrusively.
In addition, we regard all function instances as equally consuming the memory limited by the data that do not contain function-wise memory consumption. 
They actually require different memory allocations, yet the assumption is still reasonable as these functions are usually responsible for simple tasks with confined memory size~\cite{Defuse, ChallengeStudy}.
We leave these system-layer challenges to future work beyond this paper's scope.

\subsubsection{Ignorance of the function priority}
We simply regard all functions with the same level of importance or urgency, 
but a real-world platform may prioritize functions to ensure that time-sensitive or mission-critical workloads are executed promptly, even during periods of high demand or resource constraints to meet QoS requirements~\cite{Sequoia}.
We plan to improve \tool by adding a hierarchical module to dictate how to prioritize, schedule or queue functions in a QoS-aware manner.

\subsection{Threats to Validity}\label{sec:threat}
\subsubsection{Internal}
The evolution of serverless functions is very fast. With only 14-day data, we can observe distinct concept shifts in function invocations. We currently do not conduct proactive re-categorization or rule generation. The performance of \tool may degrade in the very long term.
To mitigate this threat, we design adaptive strategies to handle relatively small shifts, which can easily be extended to long-term shifts.

\subsubsection{External}
The effectiveness of \tool on other datasets is yet unknown. 
To alleviate this concern, we choose Azure Function, the most widely used dataset from real-world industry, whose representativeness has been confirmed by many studies~\cite{ServerlessWild, Defuse, FaaST, FaaSRank, Harvested}.
Moreover, it is the only publicly available industrial dataset.
We will strengthen cooperation with the industry to evaluate \tool on more datasets and prompt the implementation of \tool on a real-world FaaS system.
We also release our code to facilitate reproductivity and more experiments in the community.

\section{RELATED WORK}\label{sec:review}

Numerous efforts~\cite{Hermod,IceBreaker,Xanadu,ServerlessWild,Defuse,FaaSRank,Catalyzer,FaasCache, LCS} have focused on minimizing cold starts and memory overhead for FaaS platforms. Related studies~\cite{currentTrents,FaaSLight} classify current methods into two categories: speeding up function warming (involving renovations in the system layer) and reducing occurrences of runtime cold starts (at the application layer).
Methods of these two categories are orthogonal, so they can be combined for further cold start optimization.

This first category presents new systems to optimize infrastructure and resource management, employing techniques like load balancing among worker nodes~\cite{Hermod, IceBreaker}, snapshotting~\cite{Prebreaking, vHive}, sandbox (e.g., container) scheduling and management~\cite{Ensure, Xanadu, Catalyzer, SAND, Pagurus, Resuing, PCPM}.
One of the most popular research directions is sandbox sharing, i.e., sandboxes of the previous execution are reused for new invocations since reusing the idle sandbox incurs less delay than allocating a new one~\cite{SAND}.
A recent study, Pagurus~\cite{Pagurus}, proposed a container management scheme that allows one function's idle warm container to be forked by another function to alleviate cold starts, showing promising results.
However, implementing these methods requires significant engineering efforts, system expertise, and ongoing maintenance due to the underlying platform and sandbox modifications, which pose challenges in diverse infrastructures and platforms.

The second category mitigates cold-start occurrences~\cite{ServerlessWild, Defuse, FaasCache, FaaSRank, Ensure, LCS} through a non-intrusive way.
A primary step of function scheduling is to decide when and whether to pre-load/evict a function instance to reduce cold starts.
\cite{ServerlessWild} utilizes a small histogram to monitor function inter-invocation times, benefiting workloads with clear invocation patterns by optimizing keep-alive and pre-warming.
However, it is fully data-driven and ignores the underlying invocation patterns, leaving much room for domain-knowledge-involved invocation prediction.
Defuse~\cite{Defuse} employs dependency mining from function invocation histories to optimize the keep-alive time and pre-warming. Yet, it relies on the statistical histogram and turns to a fixed keep-alive policy for more than 32\% of the functions, delivering inadequate perfection of effectiveness.
LCS~\cite{LCS} selected the least recently warm container to reduce cold starts by keeping the containers alive for a longer period.
FaaSCache~\cite{FaasCache} innovatively establishes an equivalence between keeping functions alive and keeping objects in a cache, thereby implementing function provision based on Greedy-Dual-Size Frequency object caching. However, its fundamental idea is to use up the given resources as much as possible until they are insufficient, then evict containers to minimize cold starts. Thus, its way of optimizing resources is rigid without any predictions, failing to smoothly optimize cold starts and resource usage simultaneously. This leads to a more significant waste of memory resources.

There is improvement space for these function provisioning methods as they lack an understanding of invocation patterns and fine-grained invocation prediction.
Instead, \tool adopts horses-for-courses rule-based strategies to enhance existing approaches via accurate invocation prediction. 
\tool also proposes adaptive designs and builds up inter-function correlations, delivering better memory-economic cold-start reduction.
Other efforts target a downstream task to determine a suitable node for scheduling an individual function request. 
FaaSRank~\cite{FaaSRank} and ENSURE~\cite{Ensure} attempt to pack load on the adequate number of invokers, allowing the additional unneeded invokers to idle, so as to reduce the completion time and cold starts of functions.
Our method can be combined with these methods for finer-grained scheduling.
\section{CONCLUSION}\label{sec:conclusion}
This paper proposes the first differentiated runtime serverless function provision method for cold start optimization.
Our insight is that typical invocation patterns resulting from the underlying serverless architecture and triggers with commonness can lead to predictive invocation behaviors.
Hence, we propose \tool, which provisions functions based on categorization and invocation prediction using tailored strategies.
Our rule-based solution effectively mitigates the cold start problem by addressing the challenges of efficiency, scalability, imbalance, and evolution.
Experiments demonstrate that \tool achieves significant improvement on both sides of the trade-off: economizing expensive resources and reducing cold starts to mitigate latency.
Lastly, we make our code publicly available to facilitate future research and deployment of FaaS.
\section*{Acknowledgement}
This paper was supported by the Guangdong Key Research Program (No. 2020B010165002), the Research Grants Council of the Hong Kong Special Administrative Region, China (No. CUHK 14206921 of the General Research Fund), Shenzhen Stability Science Program, and the National Natural Science Foundation of China (No. 62202511).

\bibliographystyle{IEEEtran}
\bibliography{references}

\end{document}